\documentclass[12pt
]{article}
\usepackage{amssymb}
\def\inh{\vskip 0.075truein \noindent\hangindent=12 pt \hangafter=1}

\usepackage{amsthm}\theoremstyle{remark}
\usepackage{graphicx}
\usepackage{caption}
\usepackage{subcaption}

\newcommand{\bte}{\begin{quote}\begin{theorem}}
\newcommand{\ete}[1]{\label{#1}\end{theorem}\end{quote}}
\newcommand{\bcom}{\begin{quote}\end{quote}}
\newcommand{\bex}{\begin{quote}\begin{example}}
\newcommand{\eex}[1]{\label{#1}\end{example}\end{quote}}
\newcommand{\bcon}{\begin{quote}\begin{conclusion}}
\newcommand{\econ}[1]{\label{#1}\end{conclusion}\end{quote}}
\newcommand{\bdefi}{\begin{quote}\begin{definition}}
\newcommand{\edefi}[1]{\label{#1}\end{definition}\end{quote}}

\newcommand{\blem}{\begin{quote}\begin{lemma}}
\newcommand{\elem}[1]{\label{#1}\end{lemma}\end{quote}}

\newcommand{\bpr}{\begin{quote}\begin{problem}}
\newcommand{\epr}[1]{\label{#1}\end{problem}\end{quote}}

\newcommand{\beq}{\begin{eqnarray}}
\newcommand{\eeq}[1]{\label{#1}\end{eqnarray}}

\newcommand{\bfi}{\begin{figure}[24]}
\newcommand{\efi}[1]{\caption{\label{#1}}\end{figure}}

\newcommand{\CS}{{\cal S}}

\newcommand{\x}{&=&}

\newcommand{\bexe}{\begin{quote}\begin{exercise}\inh}
\newcommand{\eexe}[1]{\label{#1}\end{exercise}\end{quote}}

\def\CS{ Comput. Struct.}
\def\CMS{ Comput. Mater. Sci.}
\def\CoS { Compos. Struct.}

\def\IJMS{ Int. J. Mech. Sci.}

\def\IJP{ Int. J. Plasticity}
\def\IJSS{ Int. J. Solids Struct.}

\def\JBE{ J. Biomech. Eng.}
\def\JE{ J. Elasticity}

\def\JEMT{ J. Eng. Mat. Tech.}

\def\JMS{ J. Mater. Sci.}
\def\JMPS{ J. Mech. Phys. Solids}

\def\MOM{ Mech. Materials}

\def\NJP{ New J. Phys.}
\def\PMM{ PMM J. Appl. Math. Mech.}

\def\PSSA{Phys. Status Solidi A}
\def\PSSB{Phys. Status Solidi B}

\usepackage{graphics,graphicx}
\usepackage{color}
\usepackage{amsmath}
\usepackage{multirow}

\usepackage{array}
\usepackage{color}
\usepackage{colortbl}

\begin{document}

{\large
\title{Auxetic two-dimensional lattice with Poisson's Ratio arbitrarily close to $-1$.}
}

\author{Michele Brun$^{a,b}$, Luigi Cabras$^{c}$}

  \date{$^a${\em Dipartimento di Ingegneria Meccanica, Chimica e dei Materiali,} \\
{\em Universit\'a di Cagliari, Piazza d'Armi, I-09123 Cagliari, Italy} \\
$^b$ {\em Department of Mathematical Sciences} \\
{\em University of Liverpool, L69 7ZL Liverpool, Italy}\\
$^c$ {\em Dipartimento di Ingegneria Civile, Ambientale e Architettura,} \\
{\em Universit\'a di Cagliari, Piazza d'Armi, I-09123 Cagliari, Italy}}


\maketitle

\vspace{10mm}\noindent




\begin{abstract}
In this paper we propose a new lattice structure having macroscopic Poisson's ratio arbitrarily close to the stability limit -1.
We tested experimentally the effective Poisson's ratio of the micro-structured medium; the uniaxial test has been performed on a thermoplastic lattice produced with a 3d printing technology.
A theoretical analysis of the effective properties has been performed and the expression of the macroscopic constitutive properties is given in full analytical form as a function of the constitutive properties of the elements of the lattice and on the geometry of the microstructure. The analysis has been performed on three micro-geometry leading to an  isotropic behaviour for the cases of three-fold and six-fold symmetry and to a cubic behaviour for the case of four-fold symmetry.
\end{abstract}


\section{Introduction}

Auxetic materials are important in practical applications for civil and aeronautical engineering, defence e\-quip\-ments, smart sensors, filter cleaning and biomechanics and in recent years the number of patent applications and publications has increased exponentially. There is also a strong interest in the theoretical modelling and in the reformulation of several aspects of mechanics, where the interval of negative Poisson's ratio, although admissible in terms of thermodynamic stability, has often been omitted in the past. This interest must be accompanied by the design of new man-made micro-structured media that can lead to negative Poisson's ratio. The term `auxetic' comes from the Greek word `$\alpha\upsilon\xi\epsilon\sigma\iota\zeta$' (auxesis: increase, grow) and was first used by Ken Evans \cite{Evans1991} (see also \cite{EvaNkaHutRog1991}) to indicate materials having negative Poisson's  ratio, expanding in the direction perpendicular to the applied tensile stress, and contracting for perpendicular compressive stress.

The Poisson's ratio $\nu$ is an indication of the mechanical properties of a medium to deform mainly deviatorically or isotropically, as described by the ratio $K/\mu$ between the bulk and the shear modulus, ranging from a so-called `rubbery' behaviour at the upper limit of $\nu$ to a `dilatational' behaviour at the lower limit of $\nu$ \cite{Milton1992}, where for a three-dimensional isotropic medium $-1\le \nu \le 1/2$.
Rubber,  most liquids and granular solids are almost incompressible ($K/\mu\gg 1$, $\nu\rightarrow 1/2$), while examples of extremely compressible materials ($K/\mu\ll 1$, $\nu< 0$) are re-entrant foams \cite{Lakes1987,FriLakPar1988} and several molecular structures \cite{Fri1992,Bau1998,Hall2008}.

Auxetic systems perform better than classical material in a number of applications, due to their superior properties. They have been shown to provide better indentation resistance \cite{Web2008,LakElms1993,ChanEvans1998}; the material flows in the vicinity of an impact as a result of lateral contraction accompanying the longitudinal compression due to the impact loading.  Hence the auxetic material densifies in both longitudinal and transverse directions, leading to increase indentation resistance. This behavior has also been correlated to the atomic packing \cite{Rouxel2007}, which has been found to be proportional to the Poisson's ratio, and to the densification mechanism under high contact pressure \cite{Rouxel2008}.
In an isotropic material, indentation resistance is roughly proportional to the ratio $E/(1-\nu^2)$, $E$ is the Young's modulus and $\nu$ the Poisson's ratio, meaning that the resistance can be strongly increased, even with respect to an incompressible material, when the Poisson's ratio is below $-1/2$. Resistance to damage is also associated with the capacity of negative Poisson's ratio materials to distribute internal energy over a larger region as opposed to common material which, in presence of stress concentrators as point forces or geometrical singularities, accumulate internal energy in a neighborhood of the concentrator leading to possible damage of the material and consequent failure.
In this sense, auxetic materials can be applied to improve protective materials or energy absorbing materials \cite{Bezazi2008}.
Furthermore, they can be also applied as a efficient membrane filter with variable permeability \cite{Alderson2000,Raspburn2001}, fasteners \cite{Choi1991,Stavrou2005}, shape memory materials \cite{Bianchi2010} and acoustic dampeners \cite{Howell1994,Scarpa2004}.
They have the ability to form dome-shaped structures when bent \cite{Lakes1987,Burke1997}
undergoing double (synclastic) curvature, as opposed to the saddle shape (anticlastic curvature) that non-auxetic materials adopt when subject to out-of-plane bending moments.
Also, they have better acoustic and vibration properties over their conventional counterparts \cite{ChenLakes1989,Ruzzene2004,ScarpaDallocchioRuzzene2003,TeeSpadoniScarpaRuzzene2010}.

There are several natural materials that have been discovered to possess negative Poisson's ratio: iron pyrites \cite{Love1944}, arsenic and bismuth \cite{GuntonSaunders1972}, cadmium \cite{Li1976}, several cubic and face-centered cubic rare gas solids along a specific crystallographic direction \cite{Bau1998} and also biomaterials such as cow teat skin \cite{Lees1991} and load-bearing cancellous bones \cite{WilLew1982}.

Artificial auxetic structures are two and three dimensional re-entrant structures, chiral structures, rotating rigid/semi-rigid units, hard molecules, liquid crystalline polymers and
microporous polymers. Extended reviews can be found in \cite{EvansAlderson2000,Greaves2011,Prawoto2012} and here we give particular attention to the most interesting models in term of mechanics. Almost all of these models are based on a simple mechanism that is treated as a unit cell leading to a global stiffening effect.
Conceptually auxetic structure have been known since 1944 \cite{Love1944}, but the first artificial experimental samples concerning a re-entrant structures, firstly proposed by Almgren \cite{Almgren1985} and Kolpakov \cite{Kolpakov1985},  were presented  in 1987 by Lakes \cite{Lakes1987}.
Since then, different model were proposed and analyzed (see, for example, \cite{Evans1994,Choi1995,MasEvan1996,TheoStavPan1997,LarSigBou1997,GibAsh1997,Chanvans1997,Blu2005,Lu2011,DosReis2012,Rad2014}).
In the chiral structure proposed by Prall and Lakes \cite{PraLak1996}, the
basic unit is formed by connecting straight ligaments to central nodes and the auxetic effect is achieved through wrapping or unwrapping of the ligaments around the nodes in response to an applied force.
Its static and dynamic behaviour has been studied in the context of the generalized micro-polar  theory of elasticity \cite{SpaRuz2011,Liu2012,Baci2014}.
Other models, see for example \cite{Wojcie1989}, derive the auxetic behaviour by the rotation of the shapes when loaded; this kind of structure has been developed to produce the auxetic behaviour in micro- nano-structure networks by joining the rigid or semi-rigid shapes \cite{Grima2005,Grima2006,Grima2008a}; three-dimensional models in the linear  \cite{Grima2008b,Buckmann2014} and non linear \cite{Bertoldi2013} regimes were also proposed.

There are less examples of auxetic materials with continuous microstructure: the design of a family of two-dimensional, two-phase composite with Poisson's ratios arbitrarily close to -1 is given in \cite{Milton1992}. Successively, in the seminal paper \cite{MilCher1995}, it is shown that every combination of positive-definite effective constitutive tensor can be obtained  from a two-phase composite and particular attention was given to multi-rank laminates. In that paper the important concept of $n-$mode materials, indicating the number of easy modes of deformation, was also introduced; such a concept has strict analogies with the number of degrees of freedom in a mechanism of partially constrained structure, which is common in structural mechanics.

In this paper, we propose a novel lattice model with three different realizations that leads to a Poisson's ratio arbitrarily close to the thermodynamic limit corresponding to $\nu=-1$. The effect is achieved by the superposition of clockwise and anti-clockwise internal rotations leading to a macroscopic non-chiral effect.
In Section \ref{Sect01} we present experimental evidence of the negative Poisson's ratio approaching $\nu=-1$. In Section \ref{Sect02} we detail the kinematics of the mechanical system for three types of periodic lattices and in Section \ref{Sect03} we determine analytically the macroscopic constitutive properties of these structures. The dependence of the effective properties on the constitutive and geometrical parameters of the microstructure is shown in Section \ref{Sect04}, where a comparative analysis with hexagonal, triangular and square honeycombs is also performed. Final remarks in Section \ref{Sect05} conclude the paper.

\section{The lattice with Poisson's ratio close to $-1$}
\label{Sect01}

\begin{figure}[!htcb]
\centerline{
	\begin{tabular}{c}
         	 	\begin{tabular}{c@{\hspace{0.5pc}}c@{\hspace{0.5pc}}c}
             		\includegraphics[width=.3\columnwidth]{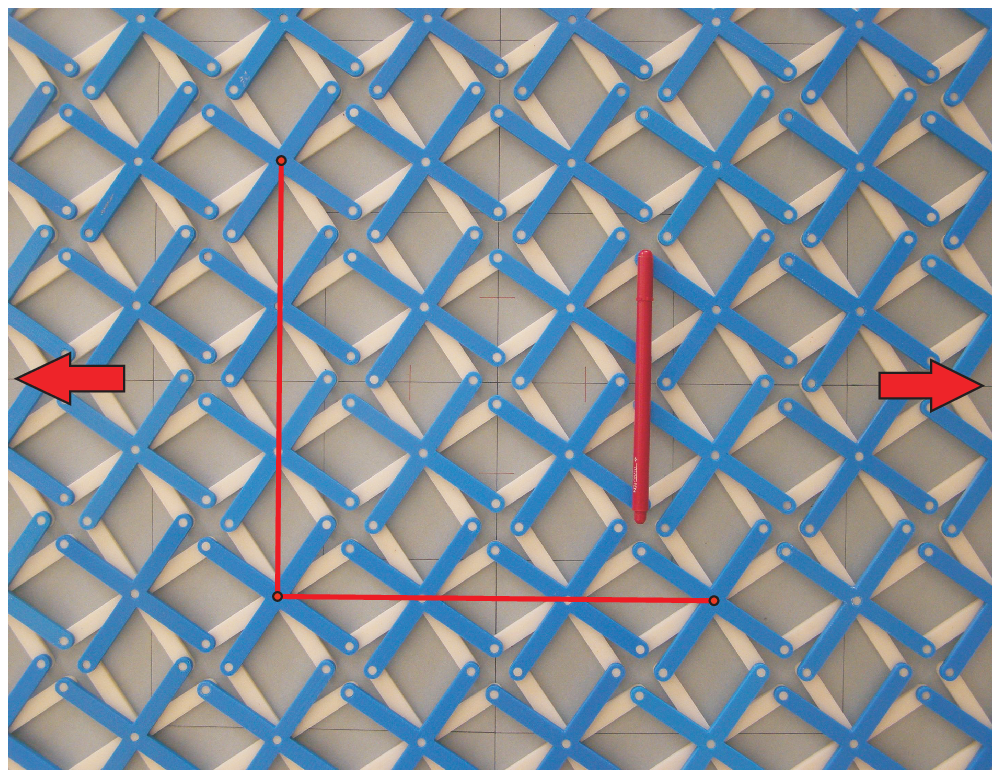} &
             		\includegraphics[width=.3\columnwidth]{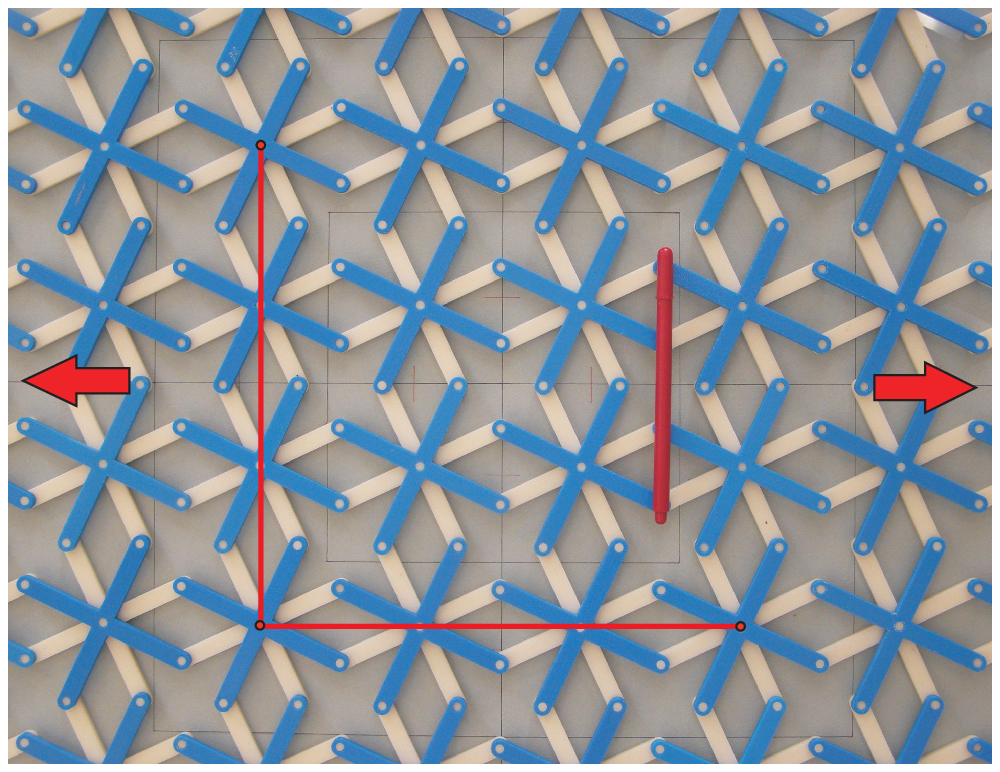} &
         			\includegraphics[width=.3\columnwidth]{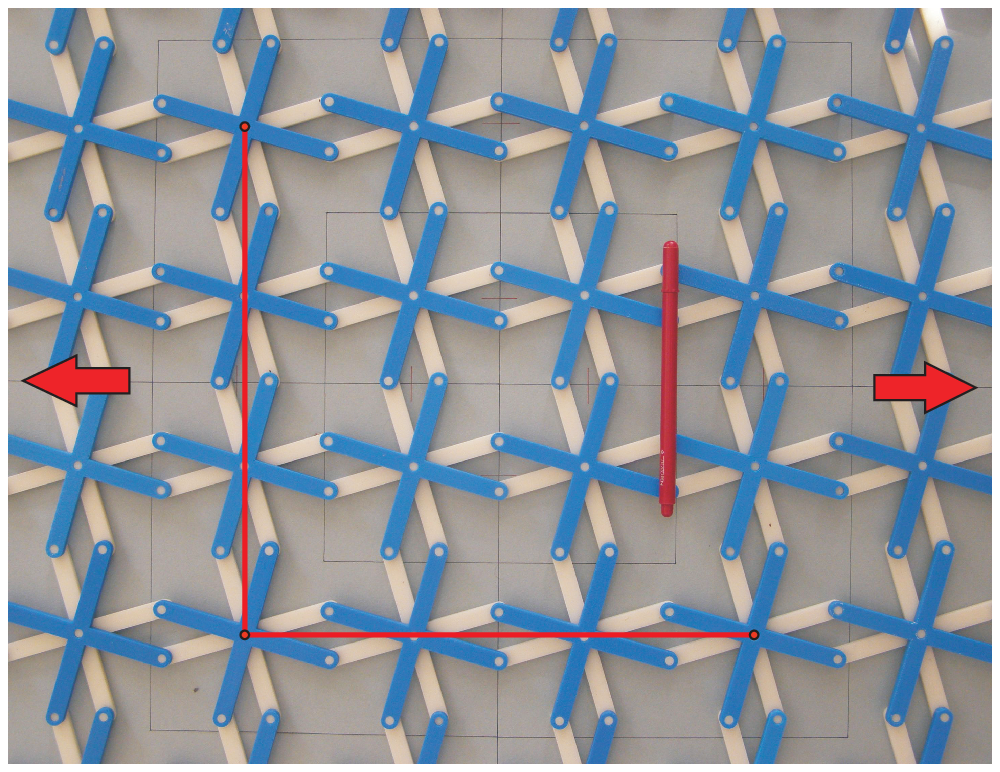}   \\
         			(a) & (b) & (c)
         		\end{tabular} \\
           	\begin{tabular}{c@{\hspace{0.6pc}}c}
         			\includegraphics[height=.4\columnwidth]{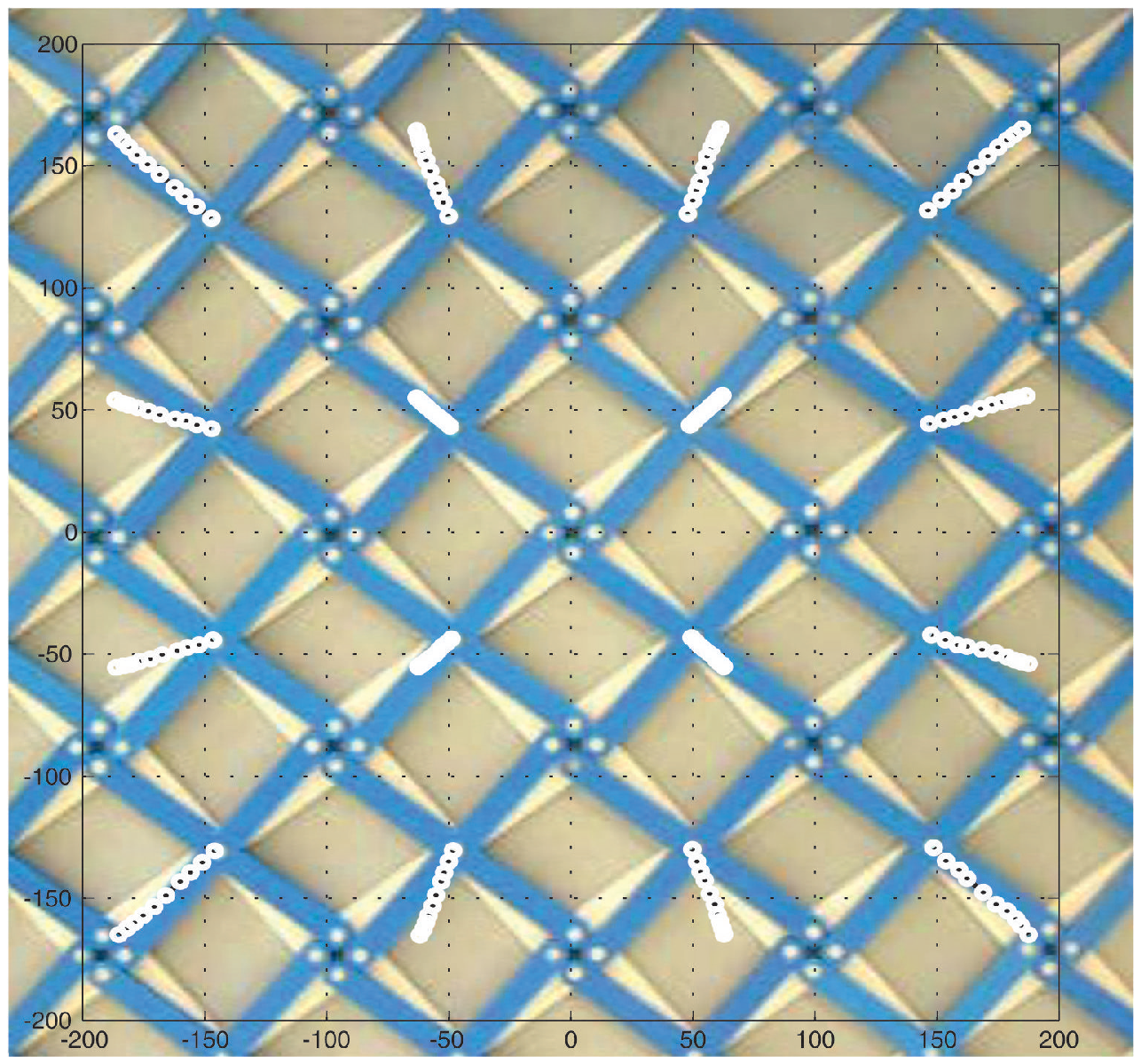} &
         			\includegraphics[height=.4\columnwidth]{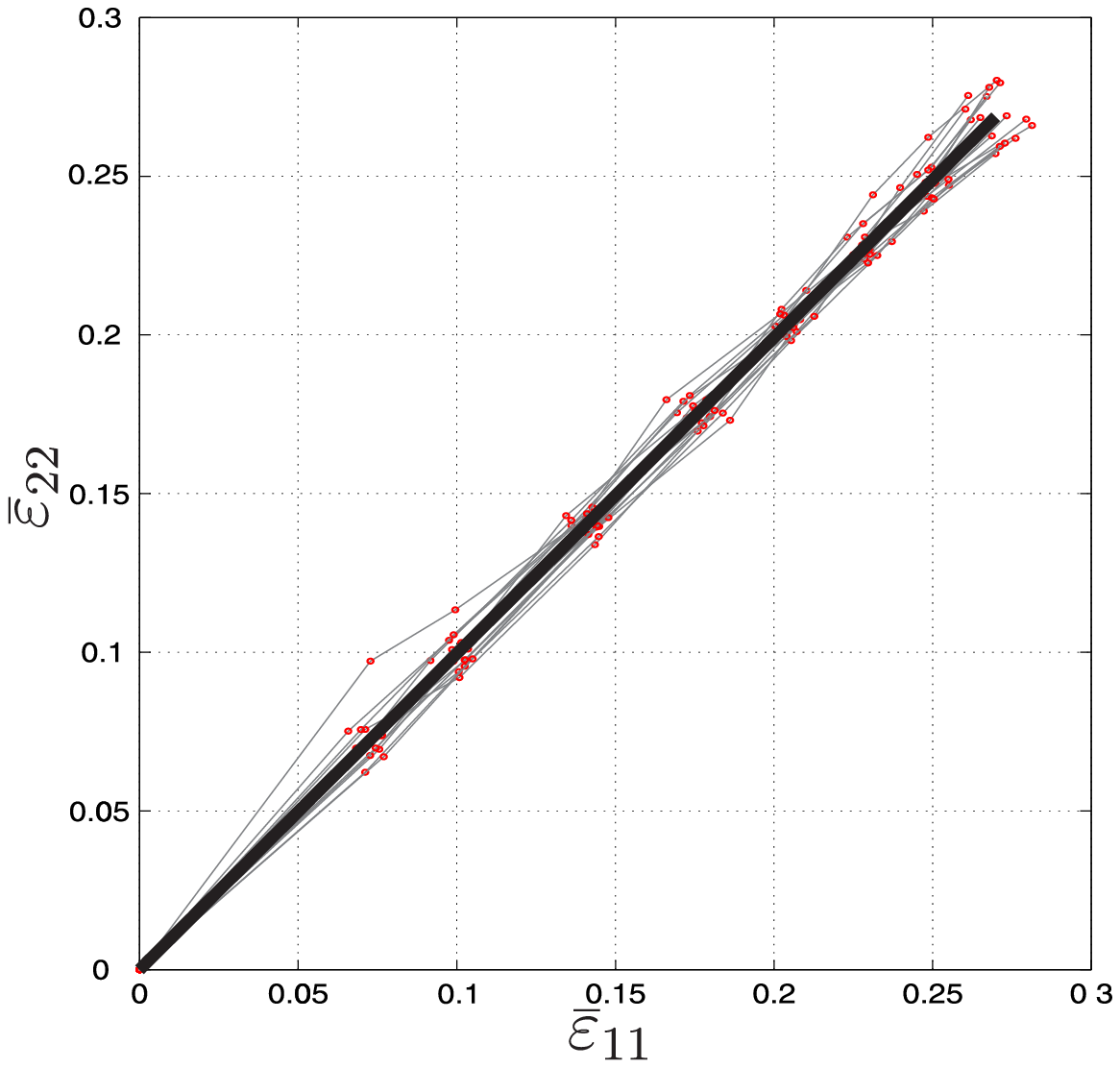} \\
         			(d) & (e)
         		\end{tabular}
         \end{tabular}
       }
\caption{\footnotesize Deformation of the auxetic lattice subjected to an horizontal tensile traction. (a), (b) and (c) show three configurations of the structured media at increasing magnitude of deformation. (d) The white dots indicate the progressive position of the central points in the $4\!\times\! 4$ central unit cells of the structured media. (e) Deformations $\overline\varepsilon_{11}$ and $\overline\varepsilon_{22}$ as a result of the applied uniaxial stress $\overline\sigma_{11}$. The grey lines correspond to the points highlighted in part (d). The thick black line indicates the average deformation. The Poisson's ratio is equal to $-0.9993$.}
\label{Fig02}
\end{figure}

Here we show experimental evidence of the micro-structured media with Poisson's ratio approaching $-1$. The elements of the structure have been produced with a 3d printer (Dimension SST 1200es) in thermoplastic polymer ABS with two different colors, blue and white.
In figure \ref{Fig02} some snapshots of the experiment are shown, the elastic structure is subjected to a uniaxial tension in horizontal direction. The images are taken at a distance of approximately $1.2$m to the sample, which has been considered sufficient to minimize image distortions.

The displacements of the junction points at the central hinge of each couple of cross-shaped elements of different colors is equal to the macroscopic displacement of the perfectly periodic structure.
The displacement of these junction points in the $4\!\times\! 4$ central unit cells are tracked from the movie of the experiment (added as supplementary material).  To this purpose an algorithm  for Image Processing in Matlab\textsuperscript{\textregistered} (Release 2011b) has been implemented and in figure \ref{Fig02}d the progressive position of these points is highlighted with white dots. The corresponding deformation is plotted in figure \ref{Fig02}e; the resulting Poisson's ratio is $\nu=-0.9993$!

\section{Model of periodic lattice with auxetic macroscopic behavior}
\label{Sect02}

The micro-structured media fall within the class of unimode materials as shown in \cite{MilCher1995}, \cite{Mil2002} (Chap. 30) and \cite{Mil2013a,Mil2013b}. In our plane linear elastic system  a single eigenvalue of the effective elasticity matrix is very small (approaching zero) and the other two are very large. As common to all isotropic and cubic materials having Poisson's ratio approaching $-1$ the only easy mode of deformation is the dilatation (plane dilatation in a two-dimensional system). Here, we focus on three affine materials, two isotropic and one cubic, presented in Section \ref{Sect01}.
The kinematic analysis of a single radially foldable structure is used to determine the Poisson's ratio of the perfect lattice and its class. Here we use the term perfect to indicate that the lattice is composed of rigid elements.

\subsection{Kinematic of radially foldable structure}

\begin{figure}[htbp]
\centering
\vspace*{10mm} \rotatebox{0}{\resizebox{!}{7.cm}{%
\includegraphics[scale=1]{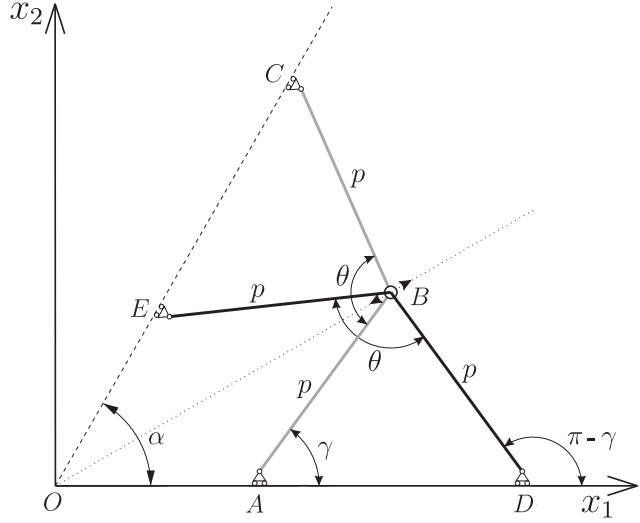}}}
\caption{Pair of linkages movable with a single degree of freedom $\gamma$.  The two rigid linkages $\overline{ABC}$ and $\overline{EBD}$ are shown in grey and black, respectively. They are constrained at the `coupler' point $B$ to have the same displacement. Points $A$ and $D$ and $E$ and $C$ can only move along straight lines.}
\label{fig2}
\end{figure}

We consider the angulated element $\overline{ABC}$, shown in grey in figure \ref{fig2}. This element represents two arms of a single cross-shaped structure that will be assembled with a second one to create the unit cell.
The rigid element $\overline{ABC}$ is supposed to roto-translate with a single degree of freedom where 
$A$ moves along the $Ox_1$-axis and $C$ moves along the axis inclined by the angle $\alpha$ with respect to the $Ox_1$-axis. In analysing the trajectory of the central point $B$, we also follow the more general formulation given in \cite{PatAna2007,YouPel1997}.
In figure \ref{fig2}, $p$ is the length of the two arms, $\theta$ the internal angle between them  and $\alpha$ is the angle between the two straight lines along which the points $A$ and $C$ are constrained to move. $B$ is the `coupler' point of the linkage.
The equation for the one-parameter trajectory followed by the point $B$ is obtained by fixing the values of the geometric variables $p$, $\theta$, $\alpha$; then, the position of $B$ is determined by the angle $\gamma$.

The coordinates of points $A$, $B$ and $C$ are (see figure \ref{fig2})
\begin{eqnarray}
\nonumber
A \equiv(x_1-p\cos\gamma,\,x_2-p\sin\gamma), \,\,\,
B\equiv(x_1,\,x_2), \\
C\equiv(x_1+p\cos(\pi-\theta+\gamma),\,x_2+p\sin(\pi-\theta+\gamma)).
\label{eqn001}
\end{eqnarray}
Then, the `coupler' point $B$ is constrained to move within the rotated ellipse
\begin{equation}
C_{1}x_1^2+C_{2}x_2^2+C_{12}x_1x_2+C=0,
\label{equa1}
\end{equation}
where
\begin{eqnarray}
\nonumber
C_{1}=p^2\,\tan^2\alpha, \\
\nonumber
C_{2}=p^2\left(2+\tan^2\alpha-2\cos\theta+2\tan\alpha\,\sin\theta \right), \\
\nonumber
C_{12}=-2p^2\left(\tan^2\alpha\,\sin\theta-\tan\alpha\,\cos\theta+\tan\alpha \right),\\
C=-p^4{\left(\tan\alpha\,\cos\theta+\sin\theta\right)}^2.
\label{equa2d}
\end{eqnarray}
Note that, for a single linkage, angles $\alpha$ and $\theta$ are independent.

When we couple the movement of the linkage $\overline{ABC}$ with the linkage $\overline{EBD}$,  depicted in black in figure \ref{fig2}, we obtain a relation between angles $\alpha$ and $\theta$.
The two linkages share the same coupler curve (\ref{equa1}) at their common point $B$ at which they are connected by means of a hinge. This implies the condition
\begin{equation}
\alpha=\pi-\theta.
\label{eqn003}
\end{equation}
Consequently, in the trajectory equation (\ref{equa1})
\begin{eqnarray}
\nonumber
C_{1}=p^2\,\tan^2\theta, \quad
C_{2}=p^2\frac{(1-\cos\theta)^2}{\cos^2\theta}, \\
C_{12}=-2p^2\left(1-\frac{1}{\cos\theta}\right)\tan\theta ,\quad
C=0,
\label{eqn004}
\end{eqnarray}
yielding
\begin{equation}
x_2=\frac{\sin\theta}{1-\cos\theta}x_1.
\label{eqn005}
\end{equation}

The common coupler curve for the two linkages is aligned with the radial line $\overline{OB}$ and to avoid crossover with other pairs in a polar arrangement of the fully radially foldable linkage the angle $\gamma$ has to satisfy the bounds
\begin{equation}
\label{equa6a}
\alpha-\eta\leq\gamma\leq\pi-\eta.
\end{equation}
where $\eta=B\widehat{A}C=B\widehat{D}E=(\pi-\theta)/2$.

\begin{figure}[htbp]
\centering
\vspace*{10mm} \rotatebox{0}{\resizebox{!}{14.5cm}{%
\includegraphics[scale=1]{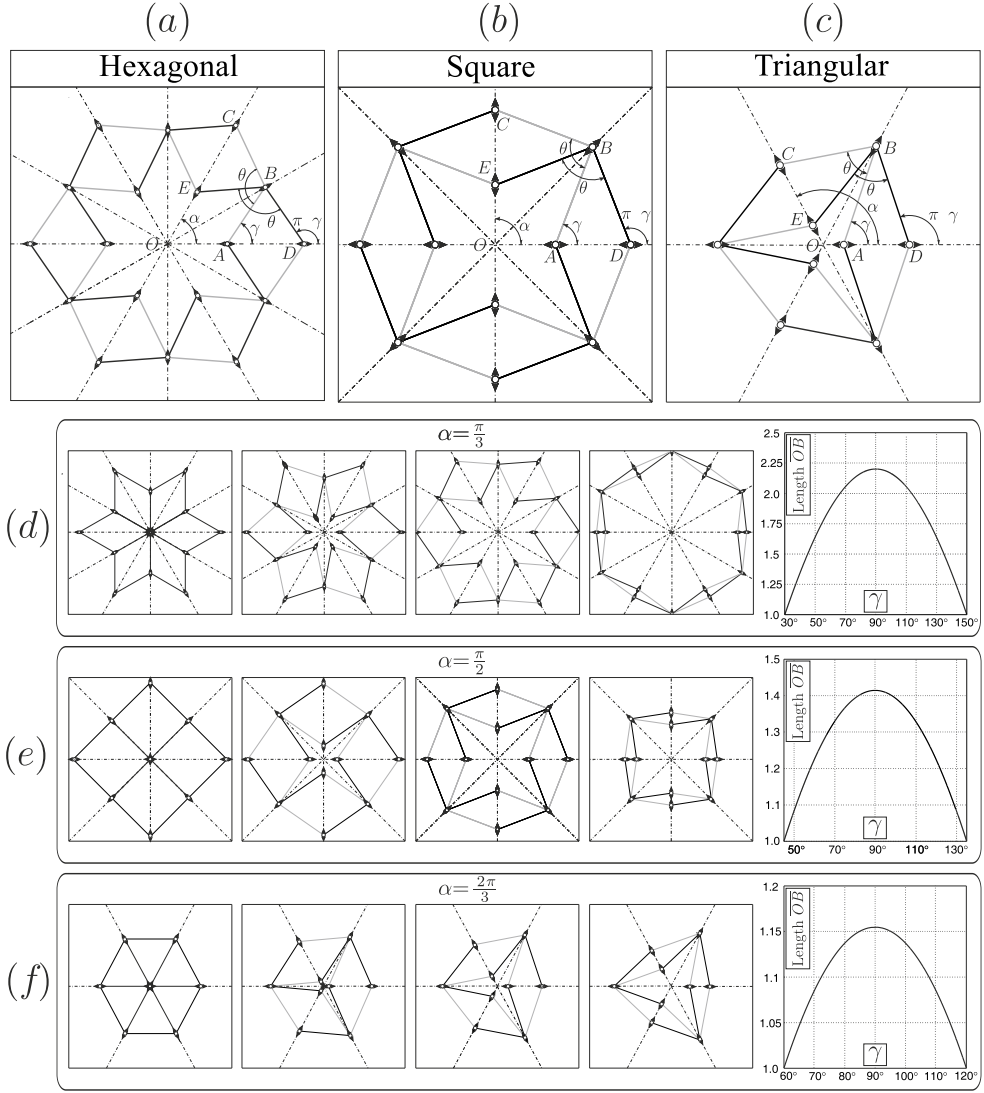}}}
 \caption{Radially foldable structures. The Poisson's ratio is equal to $-1$.
(a) $\alpha=\pi/3$, hexagonal structure composed of six couples of rigid two-arms linkages. (b) $\alpha=\pi/2$, square structure. (c) $\alpha=2\pi/3$, triangular structure.
(d-f) Configurations of the single degree of freedom unit cells at different values of the geometrical parameter $\gamma$ for the hexagonal, square and triangular structures, respectively. The radial distance $\overline{OB}$ ($p=1$) as a function of $\gamma$ is also given.}
\label{fig17}
\end{figure}

We consider three geometries: hexagonal ($\alpha=\pi/3$) figure \ref{fig17}a, square  ($\alpha=\pi/2$) figure \ref{fig17}b and triangular ($\alpha=\pi/6$) figure \ref{fig17}c. The two-arms linkages are assembled in order to create radially foldable structures. 
Different configurations are shown in figures \ref{fig17}d-f, for the hexagonal, square and triangular structures, respectively; the `coupler' point for each pair of linkages moves radially and the corresponding Poisson's ratios are equal to $-1$.

The relative position of the point $B$ with respect to the centre $O$ of the structure, shown in the right column of figure \ref{fig17}d-f as a function of the angle parameter $\gamma$, shows that the maximum volumetric expansion increases when we move from the triangular to the square and, then, to the hexagonal case.

\subsection{Construction of periodic lattice}

\begin{figure}[!htcb]
\centering
\rotatebox{0}{\resizebox{!}{11.5cm}{%
\includegraphics[scale=1]{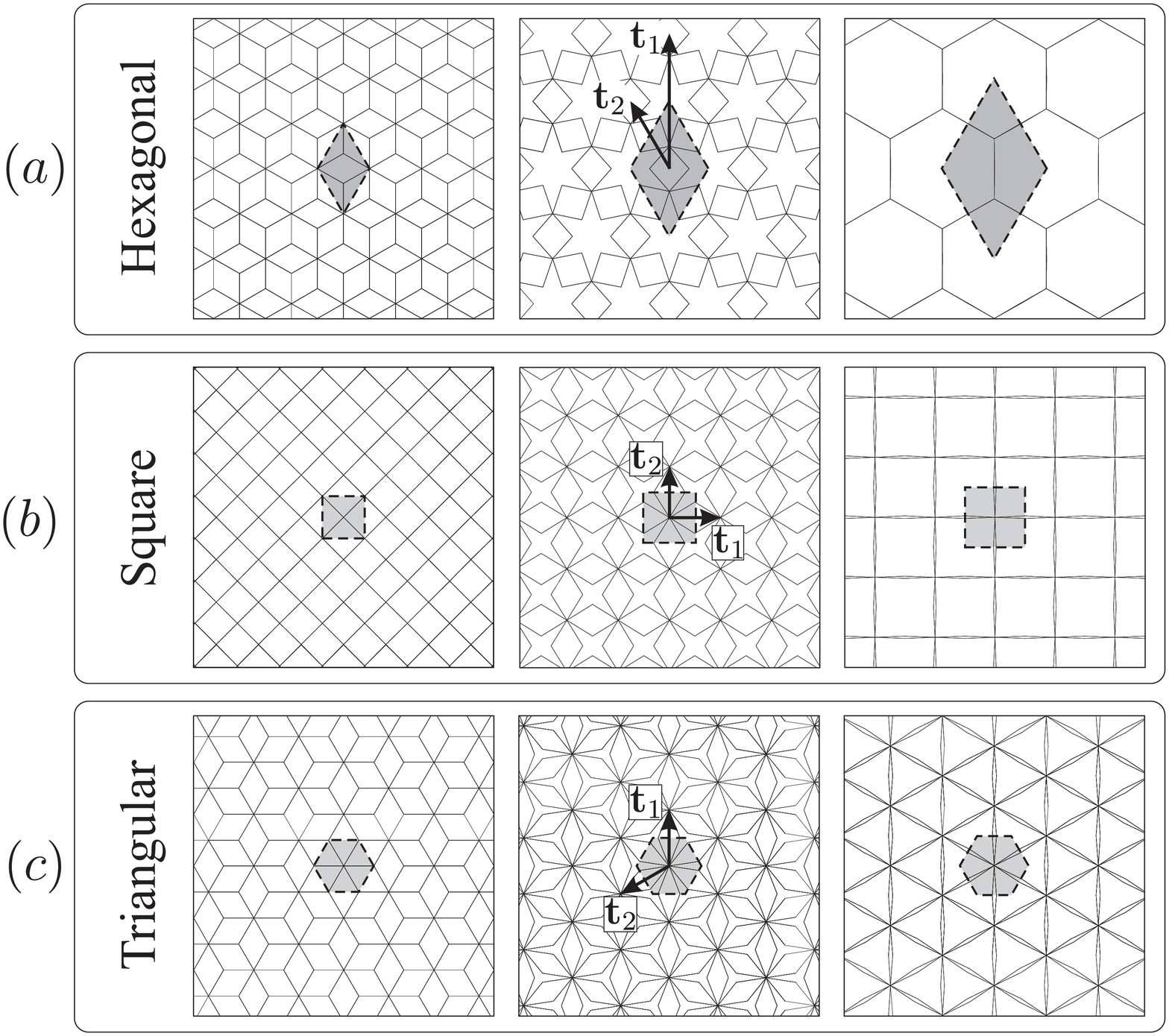}}}
 \caption{Periodic microstructures. (a) Hexagonal, (b) square and (c) triangular geometries. Three different configurations, for different values of $\gamma$, are shown for each geometry. The unit cells are also indicate The grey dashed regions are the unit cells of the Bravais lattice where ${\bf t}_1$ and ${\bf t}_2$ are the primitive vectors.}
 \label{fig5}
\end{figure}

The two-arms linkages presented in Section (\ref{Sect02}) are assembled in order to create the kinematically compatible periodic structures shown in figure \ref{fig5}. The microstructure is composed of cross-shaped elements with arms of the same length. The number of arms in each cross-shaped element are $3$, $4$ and $6$ for the hexagonal, square and triangular geometries, respectively.
A couple of cross-shaped elements is built where the two crosses are disposed in two different planes; in figure \ref{fig2} they are indicated in black and grey.
Each couple is mutually constrained to have the same displacement at the central point where a hinge is introduced. Different couples are then constrained by  internal hinges at the external end of each arm.
In the next section we will also introduce some springs to provide stability of the constitutive behaviour.

The periodic structures have a Bravais periodic lattice \cite{Kittel1996} consisting of points
\begin{equation}
{\bf R}= n_1{\bf t}_1+n_2{\bf t}_2
\label{eqn100}
\end{equation}
where $n_{1,2}$ are integers and ${\bf t}_{1,2}$ the primitive vectors spanning the lattice.
The three different geometries described in figure \ref{fig5} correspond to the fundamental centered rectangular (rhombic), square and hexagonal Bravais lattices, respectively. Following the systematic analysis for finite deformation as in \cite{Mil2013a,Mil2013b} we show here that the lattice is a unimode material. Let
\begin{equation}
{\bf T}=\left[\begin{array}{cc}
{\bf t}_1 &  {\bf t}_2
\end{array}
\right]
\label{eqn101}
\end{equation}
be the `lattice matrix'.
During the deformation the primitive vectors undergo an affine transformation and the matrix ${\bf T}$ describes a motion starting at $t=t_0$, with $\gamma(t_0)=\gamma_0$.
At time $t$ the deformation gradient is described by
\begin{equation}
{\bf F}(t,t_0)=[{\bf T}(t)][{\bf T}(t_0)]^{-1}
\label{eqn102}
\end{equation}
and the associated Cauchy-Green tensor is a path
\begin{equation}
{\bf C}(t,t_0)=[{\bf F}(t,t_0)]^T[{\bf F}(t,t_0)]=[{\bf T}(t_0)]^{-T}[{\bf T}(t)]^T[{\bf T}(t)][{\bf T}(t_0)]^{-1}
\label{eqn103}
\end{equation}
beginning at ${\bf C}(t_0,t_0)={\bf I}$.
Note that any other possible Bravais lattice is associated with the same path (\ref{eqn103}).

In particular we have the following cases.
\begin{itemize}
\item Hexagonal lattice (rhombic Bravais lattice as in figure \ref{fig5}a):
\begin{eqnarray}
{\bf T}(t)=p\,\sin\gamma \left(
                        \begin{array}{cc}
                          0 & -\sqrt{3} \\
                          6 & 3 \\
                        \end{array}
                      \right), \quad
{\bf T}(t_0)=p\,\sin\gamma_0 \left(
                        \begin{array}{cc}
                          0 & -\sqrt{3} \\
                          6 & 3 \\
                        \end{array}
                      \right)
\label{eqn104}
\end{eqnarray}
and
\begin{equation}
{\bf C}(t,t_0)=\left(\frac{\sin\gamma}{\sin\gamma_0}\right)^{2} {\bf I}.
\label{eqn105}
\end{equation}
\item Square lattice (square Bravais lattice as in figure \ref{fig5}b):
\begin{equation}
{\bf T}(t)=2\,p\,\sin\gamma\,{\bf I}, \quad
{\bf T}(t_0)=2\,p\,\sin\gamma_0\,{\bf I}
\label{eqn106}
\end{equation}
and
\begin{equation}
{\bf C}(t,t_0)=\left(\frac{\sin\gamma}{\sin\gamma_0}\right)^{2} {\bf I}.
\label{eqn107}
\end{equation}
\item Triangular lattice (hexagonal Bravais lattice as in figure \ref{fig5}c):
\begin{eqnarray}
{\bf T}(t)=2\,p\,\sin\gamma \left(
                        \begin{array}{cc}
                          0 & -\sqrt{3}/2 \\
                          1 & -1/2 \\
                        \end{array}
                      \right), \quad
{\bf T}(t_0)=2\,p\,\sin\gamma_0 \left(
                        \begin{array}{cc}
                          0 & -\sqrt{3}/2 \\
                          1 & -1/2 \\
                        \end{array}
                      \right)
\label{eqn108}
\end{eqnarray}
and
\begin{equation}
{\bf C}(t,t_0)=\left(\frac{\sin\gamma}{\sin\gamma_0}\right)^{2} {\bf I}.
\label{eqn109}
\end{equation}
\end{itemize}
Then, we conclude that the three materials are unimode, since the possible paths ${\bf C}(t,t_0)$, for any Bravais lattice, lie on the same one-dimensional curve.

\section{Effective Properties of the Periodic Auxetic Lattice}
\label{Sect03}

In this section we derive the effective constitutive parameters of the lattices. A perfect lattice would clearly have zero in-plane bulk modulus and unbounded shear moduli.
In order to estimate the macroscopic behaviour for a real lattice, we consider the elastic behaviour of the elements of the lattice and we compute the effective constitutive behaviour as a function of the constitutive behaviour of the single constituents and of the microstructure. To ensure stability we introduce extensional or rotational elastic springs that can also mimic the effect of friction in a loading branch or model elastic hinges (see figure \ref{fig6} for the hexagonal lattice).

For simplicity we restrict the attention to macroscopic linear elasticity. The linearised behaviour depends nonlinearly on the reference configuration determined by $\gamma_0$, that will be indicated as $\gamma$ in the following for ease of notation.

The hexagonal and triangular lattices have three-fold symmetries and basic considerations on the symmetry group of the material lead to the conclusion that the constitutive behaviour is isotropic (in the plane of deformation). Similar considerations, based on the tetragonal symmetry of the square lattice, lead to the conclusion that the square lattice is cubic. Therefore, it will be necessary to compute two effective elastic constants for the isotropic structures and three for the cubic one. Stability constrains the in-plane Poisson's ratio to range between $-1$ and $+1$.
Effective properties are denoted as $K^*$ (bulk modulus), $E^*$ (Young's modulus), $\mu^*$ (shear modulus) and  $\nu^*$ (Poisson's ratio) and macroscopic stress and strain as $\overline{\bf \sigma}$ and $\overline{\bf \varepsilon}$, respectively.

\subsection{Analysis of hexagonal lattice}

\begin{figure}[!htcb]
\centerline{
         \begin{tabular}{c@{\hspace{1.pc}}c}
                \includegraphics[width=4 cm]{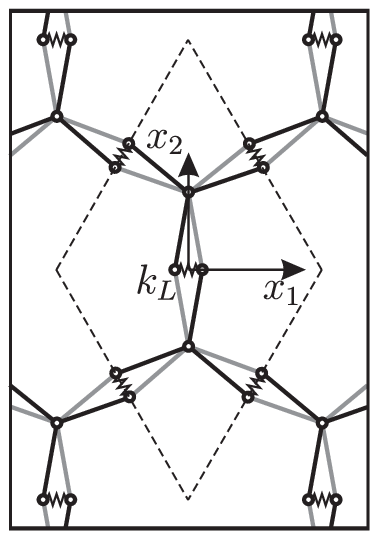} &
                \includegraphics[width=4 cm]{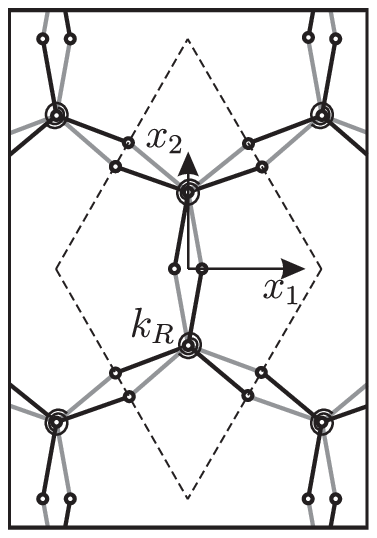}    \\
         			(a) & (b)
         \end{tabular}
}
\caption{Hexagonal lattice reinforced with elastic springs. (a) Longitudinal springs of stiffness $k_L$. (b) Rotational spring of stiffness $k_R$. The dashed area represent a typical unit cell of the periodic elastic system.}
\label{fig6}
\end{figure}

The analytical derivation of the macroscopic properties for the periodic hexagonal lattice is given.
The structure is composed of slender crosses and classical structural theories can be conveniently applied to analyse the response of the elastic system. In particular, each arm of a single cross-shaped element is modelled as an Euler beam undergoing flexural and extensional deformations. Each beam has Young's modulus $E$, cross-sectional area $A$ and second moment of inertia $J$. Additional springs have longitudinal stiffness equal to $k_L$ or rotational stiffness equal to $k_R$ (see figure \ref{fig6}).
We also introduce the non-dimensional stiffness ratio parameters $\alpha_1=k_Lp/(EA)$, $\alpha_2=k_Lp^3/(EJ)$, $\alpha_3=k_R/(pEA)$ and $\alpha_4=k_Rp/(EJ)$.
Macroscopic stresses are computed averaging the resultant forces on the boundary of the unit cell.
Periodic boundary conditions have been applied on the boundary of the unit cell so that displacements are periodic and forces are anti-periodic. Additional constraints are introduced to prevent rigid body motions.
To solve the structure we apply the Principle of Virtual Work (PVW). 
In the following, we apply the PVW in two steps: in the first we find the internal actions (bending moments $M$, shear forces $V$, axial forces $P$ and spring forces $S^L$ and moments $M^R$) of the structure searching for the kinematic admissible configuration in the set of statically admissible ones (\emph{Flexibility Method}) and in the second step we compute the macroscopic displacements. This procedure has the advantage of maintaining a sufficiently simple analytical treatment. We point out that all the results have been also verified numerically implementing a finite element code in Comsol Multiphysics\textsuperscript{\textregistered}.

\begin{figure}[!htcb]
\centerline{
         \begin{tabular}{c@{\hspace{0.5pc}}c@{\hspace{0.5pc}}c}
                \includegraphics[width=4.5 cm]{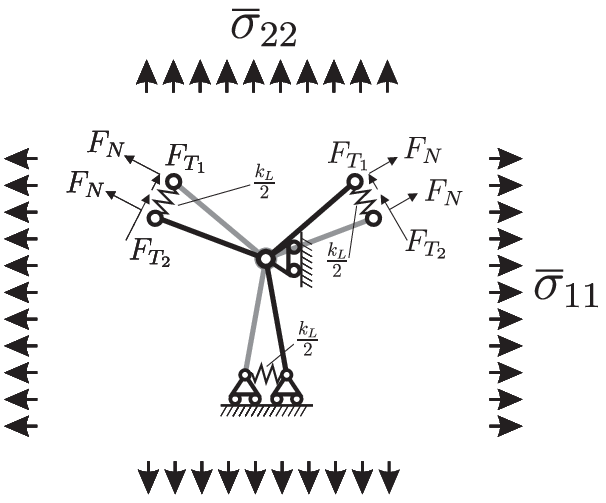} &
                \includegraphics[width=4 cm]{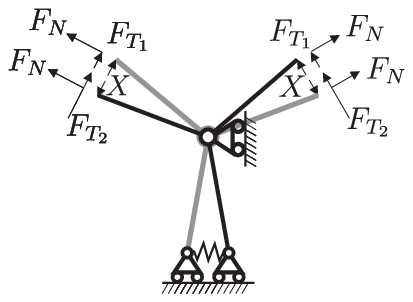} &
                \includegraphics[width=4 cm]{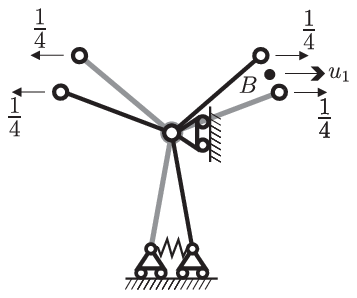} \\
         			(a) & (b) & (c)
         \end{tabular}
}
\caption{Application of the Principle of Virtual Work. Hexagonal lattice reinforced with longitudinal springs. (a) Simplified structured analysed for the computation of the effective properties. The applied forces $F_N$ and $F_{T_1}$ and $F_{T_2}$, with $F_T=(F_{T_1}+F_{T_2})/2$ correspond to macroscopic stresses components as in (\ref{eqn205}). (b) Deconnected statically determined structure introduced for the determination of the internal actions $(M,N,S_L)$. (c) Statically determined structure adopted for the computation of the horizontal displacement $u_1$ of the point $B$.}
\label{fig14}
\end{figure}

We consider the elastic structure as in figure \ref{fig14}a, subjected to known normal and tangential external forces, corresponding to a macroscopic stress having components $\overline\sigma_{11}$ and $\overline\sigma_{22}$ different from zero.
We define an equivalent statically determined (or isostatic) system disconnecting two springs and introducing the dual static parameter as unknown $X$, equal for the two springs, as shown in figure \ref{fig14}b. Then, the general field of tension $\Xi$ ($\Xi=M,V,N,S_R,M_R$) in equilibrium with the external loads is:
\begin{equation}
\Xi=\Xi_0+X\Xi_1
\label{eqn200}
\end{equation}
where $\Xi_0$ is the solution of the static scheme in equilibrium with the external loads and $X=0$; while the field $\Xi_1$ is the solution of the static scheme in equilibrium with zero external loads (autosolution of the problem) and $X=1$.

The kinematic constraints, suppressed in the isostatic structure, are restored imposing the kinematic compatibility equation that determines the values of the unknown $X$ and uniquely defines the elastic solution of the problem, i.e.,
\begin{eqnarray}
\nonumber
\displaystyle{
X=-\frac{
\sum_{beam}\int_{0}^p\left(M_0\frac{ M_1}{EJ}+N_0\frac{N_1}{EA}\right)d\xi + \sum_{spring} S^L_0 \frac{S^L_1}{k_L/2}}
{\sum_{beam}\int_0^p\left(M_1\frac{M_1}{EJ}+N_1\frac{N_1}{EA}\right)d\xi +\sum_{spring}S^L_1\frac{S^L_1}{k_L/2}}}=\\
=\left(F_N-\frac{\sqrt{3}(1+\alpha_1)}{3+3\alpha_1\cos^2\gamma+\alpha_2\sin^2\gamma}F_T\right)\cot \gamma,
\label{eqn201}
\end{eqnarray}
where $F_T=(F_{T_1}+F_{T_2})/2$.
We note that for sufficiently slender beam structures, the contribution due to the shear deformation is negligible compared to that due to flexural and axial deformations and, therefore, it has been neglected.

We reconstruct the distribution of internal actions by a linear combination of partial diagrams of $N$ and $M$
and of the spring forces  $S^L$, as functions of external forces $F_N$ and $F_T=(F_{T1}+F_{T_2})/2$:
\begin{equation}
   \begin{cases}
   N = N_0 + X N_1,\\
   M = M_0 + X M_1,\\
   S^L = S^L_0 + X S^L_1.
   \end{cases}
\label{eqn201a}
\end{equation}

Applying the PVW for the second time, we calculate the displacement of point $B$ at the centre of the spring as shown in figure \ref{fig14}c.
To this purpose we consider the real structure as kinematically admissible,  and an isostatic structure subjected to horizontal and vertical forces of magnitude equal $1/4$ as statically admissible, so that the virtual external works coincide exactly with the horizontal and vertical displacements of the point $B$, $u_1$ and $u_2$, respectively. In particular the PVW equations have the form
\begin{equation}
u_i=\sum_{beam}\int_{0}^p\left(M_i^*\frac{ M}{EJ}+N_i^*\frac{N}{EA}\right)d\xi + \sum_{spring} S^{L* }_i\frac{S^L}{k_L/2},\quad (i=1,2),
\label{eqn202}
\end{equation}
where $(M^*_i,N^*_i,S^{L*}_i)$ $(i=1,2)$ are the internal actions of the statically admissible structure subjected to forces applied in direction $x_i$ $(i=1,2)$.
The corresponding displacements are
\begin{eqnarray}
\nonumber
u_1=A_1\,F_N+B_1\,F_T, \\
u_2=A_2\,F_N+B_2\,F_T,
\label{eqn203}
\end{eqnarray}
where
\begin{eqnarray}
\nonumber
A_1=\frac{\sqrt{3}(\alpha_1+\cos^2\gamma)}{2\sin^2\gamma\,k_L},\\
\nonumber
B_1=\frac{9\cos^2\gamma(\cos^2\gamma\sin^2\gamma\!-\!1)\alpha_1^2\!-\!9\cos^2\gamma(\cos^2\gamma\!+\!1)\alpha_1\!+\!6\cos^2\gamma\sin^4\gamma\,\alpha_1\alpha_2}{6\sin^2\gamma(3+3\alpha_1\cos^2\gamma+\alpha_2\sin^2\gamma)k_L}\\
\nonumber
+\frac{\sin^6\gamma\,\alpha_2^2\!+\!3\sin^4\gamma\,\alpha_2\!-\!9\cos^4\gamma}
{6\sin^2\gamma(3+3\alpha_1\cos^2\gamma+\alpha_2\sin^2\gamma)k_L},\\
\nonumber
A_2=3\frac{3\sqrt{3}\cos^2\gamma\,\alpha_1^2+\sin^2\gamma\,\alpha_1\alpha_2+3(1+\cos^4\gamma)\alpha_1+\cos^2\gamma\sin^2\gamma\,\alpha_2+3\cos^2\gamma}
{2\sin^2\gamma(3+3\alpha_1\cos^2\gamma+\alpha_2\sin^2\gamma)k_L},\\
\nonumber
B_2=-\sqrt{3}\frac{9\cos^2\gamma(\cos^2\gamma\sin^2\gamma\!+\!1)\alpha_1^2\!+\!6\sin^2\gamma(\cos^2\gamma\sin^2\gamma\!+\!1)\alpha_1\alpha_2}{6\sin^2\gamma(3+3\alpha_1\cos^2\gamma+\alpha_2\sin^2\gamma)k_L}\\
-\frac{9(2-\cos^2\gamma\sin^2\gamma)\alpha_1+\sin^6\gamma\,\alpha_2^2\!+\!3(1-\cos^4\gamma)\alpha_2\!+\!9\cos^2\gamma}
{6\sin^2\gamma(3+3\alpha_1\cos^2\gamma+\alpha_2\sin^2\gamma)k_L}.
\label{eqn204}
\end{eqnarray}

Eqns. (\ref{eqn203}) are explicit linear relations between the forces $F_N$ and $F_T$ associated with macroscopic stresses
\begin{equation}
\overline\sigma_{11}=\frac{\sqrt{3}F_N+F_T}{3p\sin\gamma},\quad
\overline\sigma_{22}=\frac{\sqrt{3}F_N-3F_T}{3p\sin\gamma}
\label{eqn205}
\end{equation}
and displacement of the point $B$ associated with macroscopic strains
\begin{equation}
\overline\varepsilon_{11}=\frac{2u_1}{\sqrt{3}p\sin\gamma}=2\frac{A_1F_N+B_1F_T}{\sqrt{3}p\sin\gamma},\quad
\overline\varepsilon_{22}=\frac{2u_2}{3p\sin\gamma}=2\frac{A_2F_N+B_2F_T}{3p\sin\gamma}.
\label{eqn206}
\end{equation}
Solving relations (\ref{eqn206}) in terms of $F_N$ and $F_T$ and substituting the result into eqn. (\ref{eqn205}) leads to the macroscopic constitutive relation between
the macroscopic stress $\overline{\bf \sigma}$ and macroscopic strain $\overline{\bf \varepsilon}$. 
Clearly, appropriate choices of the forces $F_N$ and $F_T$ can be considered in order to set to zero some components of the stress.

The Poisson's ratio $\nu^*$ of the hexagonal lattice is:
\begin{equation}
\nu_{H_L}^*=\frac{\overline\sigma_{22}\overline\varepsilon_{11}-\overline\sigma_{11}\overline\varepsilon_{22}}{\overline\sigma_{11}\overline\varepsilon_{11}-\overline\sigma_{22}\overline\varepsilon_{22}}=
\frac{c_1\alpha_1^2+c_2\alpha_2^2+c_3\alpha_1\alpha_2+c_4\alpha_1+c_5\alpha_2+c_6}{c_7\alpha_1^2+c_2\alpha_2^2+c_8\alpha_1\alpha_2+c_9\alpha_1+c_{10}\alpha_2-c_6}
\label{equ207}
\end{equation}
where
\begin{eqnarray}
\nonumber
c_1=9\cos^2\gamma(\cos^4\gamma-\cos^2\gamma+2),\quad
c_2=-\sin^6\gamma, \quad
c_3=3\sin^2\gamma(2\cos^4\gamma-2\cos^2\gamma+1),\\
\nonumber
c_4=9(2\cos^4\gamma+\cos^2\gamma+1),\quad
c_5=-3(2\cos^4\gamma-3\cos^2\gamma+1),\quad
c_6=18\cos^2\gamma,\\
\nonumber
c_7=9\cos^2\gamma(\cos^4\gamma-\cos^2\gamma-2),\quad
c_8=3\sin^2\gamma(2\cos^4\gamma-2\cos^2\gamma-3),\\
\nonumber
c_9=-9(2\cos^4\gamma-\cos^2\gamma+3),\quad
c_{10}=-3\sin^2\gamma(2\cos^2\gamma+1).\\
\label{equ208}
\end{eqnarray}
The effective in-plane bulk modulus is
\begin{equation}\label{equa8}
K_{H_L}^*=\frac{1}{2}\frac{\overline\sigma_{11}+\overline\sigma_{22}}{\overline\varepsilon_{11}+\overline\varepsilon_{22}}=
\frac{\sqrt{3}\sin^2\gamma\,k_L}{6(\cos^2\gamma+\alpha_1)}.
\end{equation}
Consequently, the Young's modulus of the hexagonal lattice with extensional springs is
\begin{equation}
E_{H_L}^*=\frac{\overline\sigma_{11}^2-\overline\sigma_{22}^2}{\overline\sigma_{11}\overline\varepsilon_{11}-\overline\sigma_{22}\overline\varepsilon_{22}}=
2 K_{H_L}^* (1-\nu_{H_L}^*)=
\frac{4\sqrt{3} k_L  \sin^2\gamma(3+3\cos^2\gamma\,\alpha_1+\sin^2\gamma\,\alpha_2)}{-c_7\alpha_1^2-c_2\alpha_2^2-c_8\alpha_1\alpha_2-c_9\alpha_1+c_{10}\alpha_2+c_6},
\label{equ209}
\end{equation}
where the constants $c_2$, $c_{6-10}$ are given in eq. (\ref{equ208}) and the shear modulus is given by
\begin{eqnarray}
\label{equ210}
\nonumber
\mu_{H_L}^{*}=\frac{1}{2}\frac{\overline\sigma_{11}-\overline\sigma_{22}}{\overline\varepsilon_{11}-\overline\varepsilon_{22}}=
\frac{1-\nu_{H_L}^{*}}{1+\nu_{H_L}^{*}}K_{H_L}^{*}\\
=\frac{(3+3\cos^2\gamma\,\alpha_1+\sin^2\gamma\,\alpha_2)\,\sqrt{3}\,k_L}{9\cos^4\gamma\,\alpha_1^2+\sin^4\gamma\,\alpha_2^2+3(1+2\sin^2\gamma\,\cos^2\gamma)\alpha_1\alpha_2+9\,\alpha_1+3\,\alpha_2}.
\end{eqnarray}

In presence of rotational rather than longitudinal springs, the structure is statically determined and the effective constitutive parameters are the following
\begin{eqnarray}
\label{equ211}
\nonumber
\nu_{H_R}^*=\frac{3(1-2\cos^2\gamma)\alpha_3-(1-2\cos^2\gamma)\alpha_4+18\cos^2\gamma}{3(-3+2\cos^2\gamma)\alpha_3-(1+2\cos^2\gamma)\alpha_4-18\cos^2\gamma},\\
\nonumber
K_{H_R}^*=\frac{\sqrt{3}\,(k_R/p^2)}{2(3\sin^2\gamma\,\alpha_3+\cos^2\gamma\,\alpha_4+9\cos^2\gamma)},\\
\nonumber
E_{H_R}^*=\frac{8\,\sqrt{3}\,(k_R/p^2)}{3(3-2\cos^2\gamma)\alpha_3+(1+2\cos^2\gamma)\,\alpha_4+18\cos^2\gamma},\\
\mu_{H_R}^*=\frac{2\sqrt{3}\,(k_R/p^2)}{\alpha_4+3\,\alpha_3},
\end{eqnarray}
where we remember that  $\alpha_3=k_R/(EA\,p)$ and $\alpha_4=(k_R\,p)/(EJ)$.

\subsection{Analysis of the triangular lattice}

\begin{figure}[!htcb]
\centerline{
         \begin{tabular}{c@{\hspace{0.5pc}}c@{\hspace{0.5pc}}c}
                \includegraphics[width=4 cm]{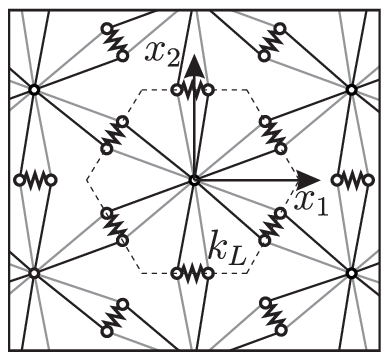} &
                \includegraphics[width=4 cm]{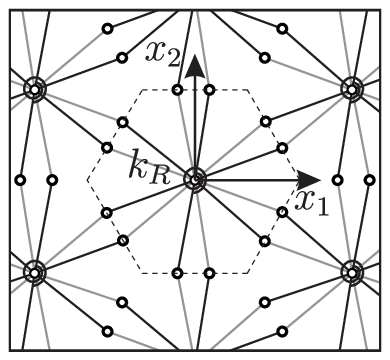} &
                \includegraphics[width=4.5 cm]{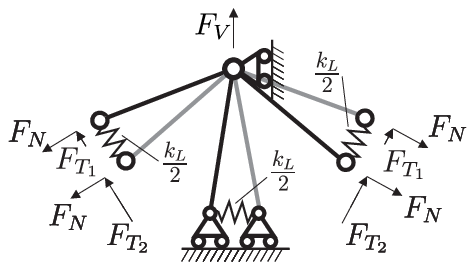} \\
         			(a) & (b) & (c)
         \end{tabular}
}
\caption{Triangular lattice reinforced with elastic springs. (a) Longitudinal springs of stiffness $k_L$. (b) Rotational spring of stiffness $k_R$. The dashed area represent a typical unit cell of the periodic elastic system. (c) Simplified structured analysed for the computation of the effective properties.}
\label{fig10}
\end{figure}

The isotropic triangular lattice structures with longitudinal and rotational springs are given in figures \ref{fig10}a and \ref{fig10}b. In part (c) of the same figure the simplified structure implemented for the computation of the macroscopic constitutive properties is also shown. The effective properties are given below.

\subsubsection{Triangular lattice with longitudinal springs}

\begin{itemize}
\item Poisson's ratio
\begin{equation}
\nu_{T_L}^*=\frac{d_1\alpha_1^3-3\alpha_1^2\alpha_2+d_2\alpha_1\alpha_2^2+d_3\alpha_1^2+d_4\alpha_2^2+d_5\alpha_1\alpha_2+d_6\alpha_1+d_7\alpha_2}
{-d_1\alpha_1^3+d_8\alpha_1^2\alpha_2+3d_2\alpha_1\alpha_2^2+d_9\alpha_1^2+3d_4\alpha_2^2+d_{10}\alpha_1\alpha_2-d_6\alpha_1-d_7\alpha_2}
\label{equ301}
\end{equation}
where
\begin{eqnarray}
\nonumber
d_1=-9\cos^4\gamma,\quad
d_2=\sin^4\gamma, \quad
d_3=9(\cos^6\gamma-4\cos^4\gamma+2\cos^2\gamma-1),\\
\nonumber
d_4=\sin^4\gamma\cos^2\gamma,\quad
d_5=3(6\cos^4\gamma-7\cos^2\gamma-2\cos^6\gamma+1),\quad
d_6=-9\cos^2\gamma,\\
\nonumber
d_7=-3\cos^2\gamma,\quad
d_9=9(3\cos^6\gamma-4\cos^4\gamma+2\cos^2\gamma+1),\\
\nonumber
d_8=3(4\sin^2\!\gamma\cos^2\!\gamma\!+1),\quad
d_{10}=3(10\cos^4\gamma-6\cos^6\gamma-5\cos^2\gamma+3).\\
\label{equ302}
\end{eqnarray}
\item  Bulk modulus
\begin{equation}
\label{equ303}
K_{T_L}^*=\frac{\sqrt{3}\sin^2\gamma \,k_L}{2\,(\cos^2\gamma+\alpha_1)}.
\end{equation}
\item Young's modulus
\begin{equation}
E_{T_L}^*=\frac{2\,\sqrt{3}\left[9\cos^4\gamma\,\alpha_1^2+\sin^4\gamma\,\alpha_2^2+3(2\cos^2\gamma\sin^2\gamma+1)\alpha_1\alpha_2+9\alpha_1+3\alpha_2\right]\sin^2\gamma\,k_L}
{-d_1\alpha_1^3+d_8\alpha_1^2\alpha_2+3d_2\alpha_1\alpha_2^2+d_9\alpha_1^2+3d_4\alpha_2^2+d_{10}\alpha_1\alpha_2-d_6\alpha_1-d_7\alpha_2},
\label{equ304}
\end{equation}
where $d_1$, $d_2$, $d_4$, $d_{6-10}$ are given in (\ref{equ302}).
\item Shear modulus
\begin{equation}
\mu_{T_L}^*=\frac{\sqrt{3}\left[9\cos^4\gamma\alpha_1^2+\sin^4\gamma\alpha_2^2+3(2\cos^2\gamma\sin^2\gamma+1)\alpha_1\alpha_2+9\alpha_1+3\alpha_2\right]k_L}
{12\cos^2\gamma\,\alpha_1^2\alpha_2\!+\!4\sin^2\gamma\,\alpha_1\alpha_2^2\!+\!9\sin^22\gamma\,\alpha_1^2\!+\!6(2\!-\!\sin^22\gamma)\alpha_1\alpha_2\!+\!\sin^22\gamma\,\alpha_2^2}.
\label{equ305}
\end{equation}
\end{itemize}

\subsubsection{Triangular lattice with rotational springs}

When rotational springs are considered, the triangular lattice structure is statically determined and the effective constants are as follows.
\begin{itemize}
\item Poisson's ratio
\begin{equation}
\nu_{T_R}^*=\frac{e_1\alpha_3^2+e_2\alpha_4^2+e_3\alpha_3\alpha_4+e_4\alpha_3+e_5\alpha_4}{e_6\alpha_3^2+e_7\alpha_4^2+e_8\alpha_3\alpha_4-e_4\alpha_3-e_5\alpha_4},
\label{equ306}
\end{equation}
where
\begin{eqnarray}
\nonumber
e_1=9\,(2\cos^4\gamma-3\cos^2\gamma+1),\quad
e_2=(2\cos^4\gamma-\cos^2\gamma), \quad
e_3=3(4\cos^2\gamma\sin^2\gamma-1),\\
\nonumber
e_4=27\cos^2\gamma,\quad
e_5=9\cos^2\gamma,\quad
e_6=9\,(2\cos^4\gamma-\cos^2\gamma-1),\\
\nonumber
e_7=2\cos^4\gamma-3\cos^2\gamma,\quad
e_8=12\cos^2\gamma\sin^2\gamma-9. \\
\label{equ307}
\end{eqnarray}
\item Bulk modulus
\begin{equation}
\label{equ308}
K_{T_R}^*=\frac{3\,\sqrt{3}\,(k_R/p^2)}{2(3\,\sin^2\gamma\,\alpha_3+\cos^2\gamma\,\alpha_4+9\cos^2\gamma)}.
\end{equation}
\item Young's modulus
\begin{equation}
\label{equ309}
E_{T_R}^*=-\frac{6\sqrt{3}\,(3\alpha_3+\alpha_4)\,(k_R/p^2)}
{e_6\alpha_3^2+e_7\alpha_4^2+e_8\alpha_3\alpha_4-e_4\alpha_3-e_5\alpha_4},
\end{equation}
where the coefficients $e_{4-8}$ are given in eq. (\ref{equ307}).
\item Shear modulus
\begin{equation}
\label{equ310}
\mu_{T_R}^{*}=\frac{3\,\sqrt{3}\,(3\,\alpha_3+\alpha_4)\,(k_R/p^2)}{9\sin^22\gamma\,\alpha_3^2+\sin^22\gamma\,\alpha_4^2+6(2-\sin^2 2\gamma)\alpha_3\alpha_4}.
\end{equation}
\end{itemize}

\subsection{Analysis of the square lattice}

\begin{figure}[!htcb]
\centerline{
         \begin{tabular}{c@{\hspace{0.5pc}}c@{\hspace{0.5pc}}c}
                \includegraphics[width=4.5 cm]{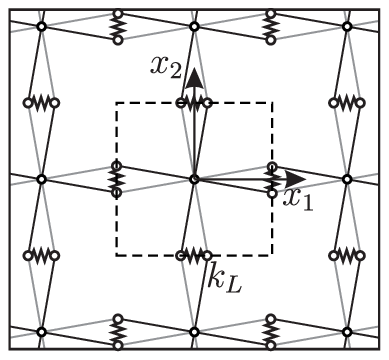}&
                \includegraphics[width=4. cm]{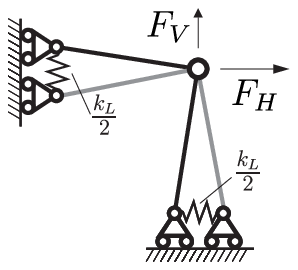}&
                \includegraphics[width=4. cm]{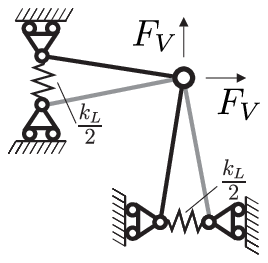} \\
         			(a) & (b) & (c)
         \end{tabular}
}
\caption{Square lattice reinforced with longitudinal elastic springs. (a) The lattice structure. The dashed area represents a typical unit cell of the periodic elastic system. (b) Simplified structure used for the computation of the effective in-plane Poisson's ratio $\nu_{S_L}^*$, bulk modulus $K_{S_L}^*$ and Young's modulus $E_{S_L}^*$. The forces $F_{H,V}$ are associated with macroscopic stress components $\overline\sigma_{11,22}=F_{H,V}/(2p\sin\gamma)$. (c) Simplified structure used for the computation of the in-plane shear modulus $\mu_{S_L}^*$. The force $F_V$ is associated with the macroscopic stress $\overline\sigma_{12}=F_{V}/(2p\sin\gamma)$. }
\label{fig11}
\end{figure}

The square lattice has cubic symmetry and it is necessary to compute three independent elastic constants to determine the effective behaviour of the structure. The lattice with longitudinal springs is given in figure \ref{fig11}a, and the simplified structure used to compute the in-plane Poisson's ratio $\nu_{S_L}^*$, bulk modulus $K_{S_L}^*$ or Young's modulus $E_{S_L}^*$ is shown in figure \ref{fig11}b. Statically the structure is twice overdetermined and it is therefore necessary to introduce two disconnections and two dual static variables to find the internal actions within the elastic system. The simplified structure introduced to compute the in-plane shear modulus $\mu_{S_L}^*$ is given in figure \ref{fig11}c. In this case the springs are not activated and they can be neglected, so that the structure can be considered as statically determined. The same structural models of figure \ref{fig11} have been considered with rotational springs instead of longitudinal for the second case. The effective properties are reported in the following.

\subsubsection{Square lattice with longitudinal springs}

\begin{itemize}
\item Poisson's ratio
\begin{equation}
\nu_{S_L}^*=\frac{-3\,(\alpha_1+1)^2\,\cos^2\gamma}
{3\cos^2\gamma\,\alpha_1^2\!+\!2\sin^2\gamma\,\alpha_1\alpha_2\!+\!6(1\!-\!\sin^2\gamma\cos^2\gamma)\alpha_1\!+\!2\sin^2\gamma\cos^2\gamma\,\alpha_2\!+\!3\cos^2\gamma}.
\label{equ401}
\end{equation}
\item  Bulk modulus
\begin{equation}
\label{equ402}
K_{S_L}^*=\frac{k_L\,\sin^2\gamma}{2\,(\alpha_1+\cos^2\gamma)}.
\end{equation}
\item Young's modulus
\begin{equation}
E_{S_L}^*=\frac{2\,k_L\sin^2\gamma(3\cos^2\gamma\,\alpha_1\!+\!\sin^2\gamma\,\alpha_2\!+\!3)}
{3\cos^2\gamma\,\alpha_1^2\!+\!2\sin^2\gamma\,\alpha_1\alpha_2\!+\!6(1\!-\!\sin^2\gamma\cos^2\gamma)\alpha_1\!+\!2\sin^2\gamma\cos^2\gamma\,\alpha_2\!+\!3\cos^2\gamma}.
\label{equ403}
\end{equation}
\item Shear modulus
\begin{equation}
\mu_{S_L}^*=\frac{3 k_L}{2(3\cos^2\gamma\,\alpha_1+\sin^2\gamma\,\alpha_2)}.
\label{equ404}
\end{equation}
\end{itemize}

\subsubsection{Square lattice with rotational springs}

The structure is once overdetermined. The effective constitutive parameters are as follows.
\begin{itemize}
\item Poisson's ratio
\begin{equation}
\nu_{S_R}^*=-\frac{3\,\cos^2\gamma}{3\sin^2\gamma\,\alpha_3+\cos^2\gamma\,\alpha_4+3\cos^2\gamma}.
\label{equ405}
\end{equation}
\item  Bulk modulus
\begin{equation}
\label{equ406}
K_{S_R}^*=-\frac{3\,(k_R/p^2)}
{2(3\sin^2\gamma\,\alpha_3+\cos^2\gamma\,\alpha_4+6\cos^2\gamma)}.
\end{equation}
\item Young's modulus
\begin{equation}
E_{S_R}^*=\frac{3\,(k_R/p^2)}
{3\sin^2\gamma\,\alpha_3+\cos^2\gamma\,\alpha_4+3\cos^2\gamma}.
\label{equ407}
\end{equation}
\item Shear modulus
\begin{equation}
\mu_{S_R}^*=\frac{3 (k_R/p^2)}{2(3\cos^2\gamma\,\alpha_3+\sin^2\gamma\,\alpha_4)}.
\label{equ408}
\end{equation}
\end{itemize}

\section{Analysis of effective properties}
\label{Sect04}

Here, the effective properties of the micro-stuctured media are analysed in detail.
We consider at first the case of vanishing stiffness of the springs $k_L,k_R\rightarrow 0$.
For every lattice
\begin{eqnarray}
\nonumber
\nu^*\simeq -1+\nu^*_1 \, k_{L,R}+\mathcal{O}\left(k_{L,R}^2\right), \quad
K^*\simeq 0+K^*_1 \, k_{L,R}+\mathcal{O}\left(k_{L,R}^2\right), \\
E^*\simeq 0+E^*_1 \, k_{L,R}+\mathcal{O}\left(k_{L,R}^2\right), \quad
\mu^*\simeq\mu^*_0+\mu^*_1\,  k_{L,R}+\mathcal{O}\left(k_{L,R}^2\right).
\label{equ501}
\end{eqnarray}
where $\nu^*_1,K^*_1,E^*_1,\mu^*_{0,1}>0$ and their explicit expressions are given in Table \ref{Table01}.
\begin {table}[!hcbt]
\begin{center}
\renewcommand\arraystretch{1.35}
\begin{tabular}{|c|c|c|}
\hline
\textbf{Lattice} & \textbf{Longitudinal springs}& \textbf{Rotational springs}\\ \hline
\multirow{4}{*}{\rotatebox{90}{Hexagonal}}&$\nu_1^*=\frac{p}{3}\frac{\eta_1}{EJ\,EA}$ & $\nu_1^*=\frac{1}{9}\frac{3\,EJ+p^2\,EA}{p\,\cos^2\gamma\,EJ\,EA}$\\
 &$K_1^*=\frac{\sqrt{3}}{6}\,\tan^2\gamma$    & $K_1^*=\frac{\sqrt{3}}{18\,p^2\cos^2\gamma}$\\
 &$E_1^*=\frac{2\,\sqrt{3}}{3}\,\tan^2\gamma$ & $E_1^*=\frac{2\,\sqrt{3}}{9\,p^2\cos^2\gamma}$ \\
 &$\mu_0^*=\frac{\sqrt{3}\,EJ\,EA}{p\,\eta_1}$ & $\mu_0^*=\frac{\sqrt{3}\,EJ\,EA}{p\,\eta_1}$\\[1 mm] \hline
 \multirow{4}{*}{\rotatebox{90}{Triangular}}&$\nu_1^*=\frac{4p}{3}\frac{\eta_2\,\eta_3}{EA\,EJ\,\eta_1}\tan^2\gamma $ & $\nu_1^*=\frac{4}{9}\frac{\eta_2\,\eta_3}{p\,\cos^2\gamma\,EJ\,EA\,\eta_1}$\\
 &$K_1^*=\frac{\sqrt{3}}{2}\,\tan^2\gamma$ & $ K_1^*=\frac{\sqrt{3}}{6\,p^2\,\cos^2\gamma}$\\
 &$E_1^*=2\,\sqrt{3}\,\tan^2\gamma$ & $E_1^*=\frac{2\,\sqrt{3}}{3\,p^2\,\cos^2\gamma}$ \\
 &$\mu_0^*=\frac{3\sqrt{3}}{4\,p}\frac{\eta_1\,EA\,EJ}{\eta_2\,\eta_3}$ & $\mu_0^*=\frac{3\sqrt{3}}{4\,p}\frac{\eta_1\,EA\,EJ}{\eta_2\,\eta_3}$ \\[1 mm]  \hline
 \multirow{4}{*}{\rotatebox{90}{Square}}&$\nu_1^*=\frac{2}{3}\frac{p\,\sin^2\gamma\,\eta_2}{EA\,EJ\,\cos^2\gamma}$ & $\nu_1^*=\frac{1}{3}\frac{\eta_2}{EA\,EJ\,p\,\cos^2\gamma}$ \\
 &$K_1^*=\frac{1}{2}\,\tan^2\gamma$ & $K_1^*=\frac{1}{4\,p^2\,cos^2\gamma}$\\
 &$E_1^*=2\,\tan^2\gamma$ & $E_1^*=\frac{1}{p^2\,cos^2\gamma}$ \\
 &$\mu_0^*=\frac{3}{2\,p}\frac{EJ\,EA}{\eta_3}$ & $\mu_0^*=\frac{3}{2\,p}\frac{EJ\,EA}{\eta_3}$\\[1 mm]   \hline
\end{tabular}
\caption {\footnotesize{Explicit expression of the coefficients in the asymptotic formulae  in eqn. (\ref{equ501}). In the table $\eta_1=3\,EJ+p^2\,EA$,\quad$\eta_2=3\,EJ\,\sin^2\gamma+p^2\,EA\,\cos^2\gamma$,\quad$\eta_3=3\,EJ\,\cos^2\gamma+p^2\,EA\,\sin^2\gamma$.}}
\label{Table01}
\end{center}
\end {table}

\begin{figure}[!htcb]
\centerline{
         \begin{tabular}{c@{\hspace{0.5pc}}c}
                \includegraphics[width=6 cm]{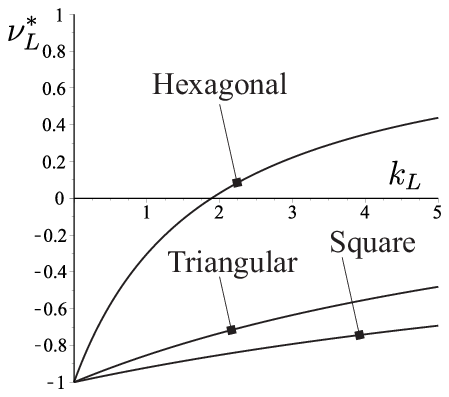} &
                \includegraphics[width=6 cm]{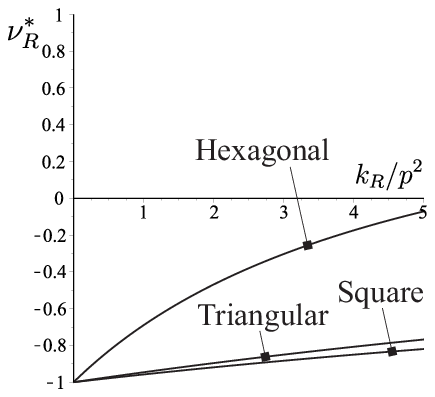}    \\
         			(a) & (b)
         \end{tabular}
}
\caption{Effective Poisson's ratio $\nu^*$ as a function of the spring stiffnsses $k_L$ and $k_R$. (a) $\nu^*_{H_L}$, $\nu^*_{T_L}$, $\nu^*_{S_L}$ are given as a function of $k_L$. (b) $\nu^*_{H_R}$, $\nu^*_{T_R}$, $\nu^*_{S_R}$ are given as a function of $k_R/p^2$. Results correspond to the following parameters: Young's modulus $E=3000$ MPa (thermoplastic polymer ABS), $A=50$ mm$^2$, $J=417$ mm$^4$ corresponding to the geometry given in Figure \ref{Fig02}.}
\label{fig12}
\end{figure}

It is shown in (\ref{equ501}) that also for deformable structures the Poisson's ratio remains $-1$ when the spring stiffnesses are zero, while the effect of the springs is to increase the value of $\nu^*$.  In such a limit, the bulk and the Young's moduli vanish while the shear modulus remains finite. This is clearly associated with the deformation mechanism of the lattice, which involve deformation of the cross-shaped elements when a macroscopic shear stress or shear deformation is applied, while macroscopic volumetric deformations can be sustained by rigid internal rotations of the elements of the microstructure.
The limiting behaviour described in (\ref{equ501}) can also be understood in terms of relative stiffness between the spring elements and the elements of the lattice as described by the coefficients $\alpha_{1,\cdots,4}$. In this respect, when $\alpha_{1,\cdots,4}\rightarrow 0$, the same outcomes of eqns.
(\ref{equ501}) are obtained.
The dependence of the Poisson's ratio on the stiffnesses $k_L$ and $(k_R/p^2)$ is shown in figure \ref{fig12}a and figure \ref{fig12}b, respectively, for the three microstructures. Results confirm that the Poisson's ratio approaches $-1$ when the spring constants are zero. They also show that $\nu^*_{90}<\nu^*_{120}<\nu^*_{60}$.

It is worthwhile to note the maximum theoretical values that can be reached by the Poisson's ratios at the limit $k_{L},k_R/p^2\rightarrow \infty$; the limiting expressions are
\begin{eqnarray}
\nonumber
\nu^*_{H_L}\simeq 1-\frac{\left(s/p\right)^2}{\sin^2\gamma}+\mathcal{O}\left(\left(s/p\right)^4\right),&
\displaystyle{\nu^*_{H_R}\simeq \frac{1\!-\!2\cos^2\gamma}{1\!+\!2\cos^2\gamma}\!\left[1-\!\frac{\left(s/p\right)^2}{1\!+\!2\cos^2\gamma}\right]\!+\!\mathcal{O}\left(\left(s/p\right)^4\right),}\\
\nonumber
\nu^*_{T_L}\simeq \frac{1}{3}\!-\!\frac{1\!+\!\sin^2\!\gamma\cos^2\!\gamma}{\sin^2\gamma}\left(s/p\right)^2\!+\!\mathcal{O}\left(\left(s/p\right)^4\right),&
\displaystyle{\nu^*_{T_R}\simeq \frac{1\!-\!2\cos^2\gamma}{3\!-\!2\cos^2\gamma}\!\left[1-\!\frac{\left(s/p\right)^2}{3\!-\!2\cos^2\gamma}\right]\!+\!\mathcal{O}\left(\left(s/p\right)^4\right)}\\
\nu^*_{S_L}\simeq 0 -\frac{\cot^2\gamma}{8}\left(s/p\right)^2+\mathcal{O}\left(\left(s/p\right)^4\right),&
\nu^*_{S_R}\rightarrow 0,
\label{equ503}
\end{eqnarray}
where in (\ref{equ503}) and in the following we consider, for simplicity, rectangular cross-sections of the arms of the cross-shaped elements, so that $A=t\,s$ and $J=t\,s^3/12$, where $s$ and $t$ are the in-plane and out-of-plane thicknesses, respectively. Therefore $s/p\ll 1$. We note that the Poisson's ratio remains always negative for the square lattice approaching zero in the limit. Interestingly, the hexagonal lattice with longitudinal springs has a completely different behaviour approaching the upper limit for the Poisson's ratio corresponding to an incompressible material. For the structures with rotational springs the limit depends on the actual configuration described by the angle $\gamma$ and ranges between $-0.2$ and $1$ for the hexagonal lattice and between $0.2$ and $1$ for the triangular one. For general values of $k_R$ and $k_L$, we note that the shear modulus $\mu^*$ is independent of the rotational stiffness $k_R$. Also, for the square lattice, the shear modulus does not depend  on the longitudinal stiffness $k_L$; in fact, $\mu_{S_L}=\mu_{S_R}$.

\begin{figure}[!htcb]
\centerline{
         \begin{tabular}{c@{\hspace{0.5pc}}c}
                \includegraphics[width=6.6 cm]{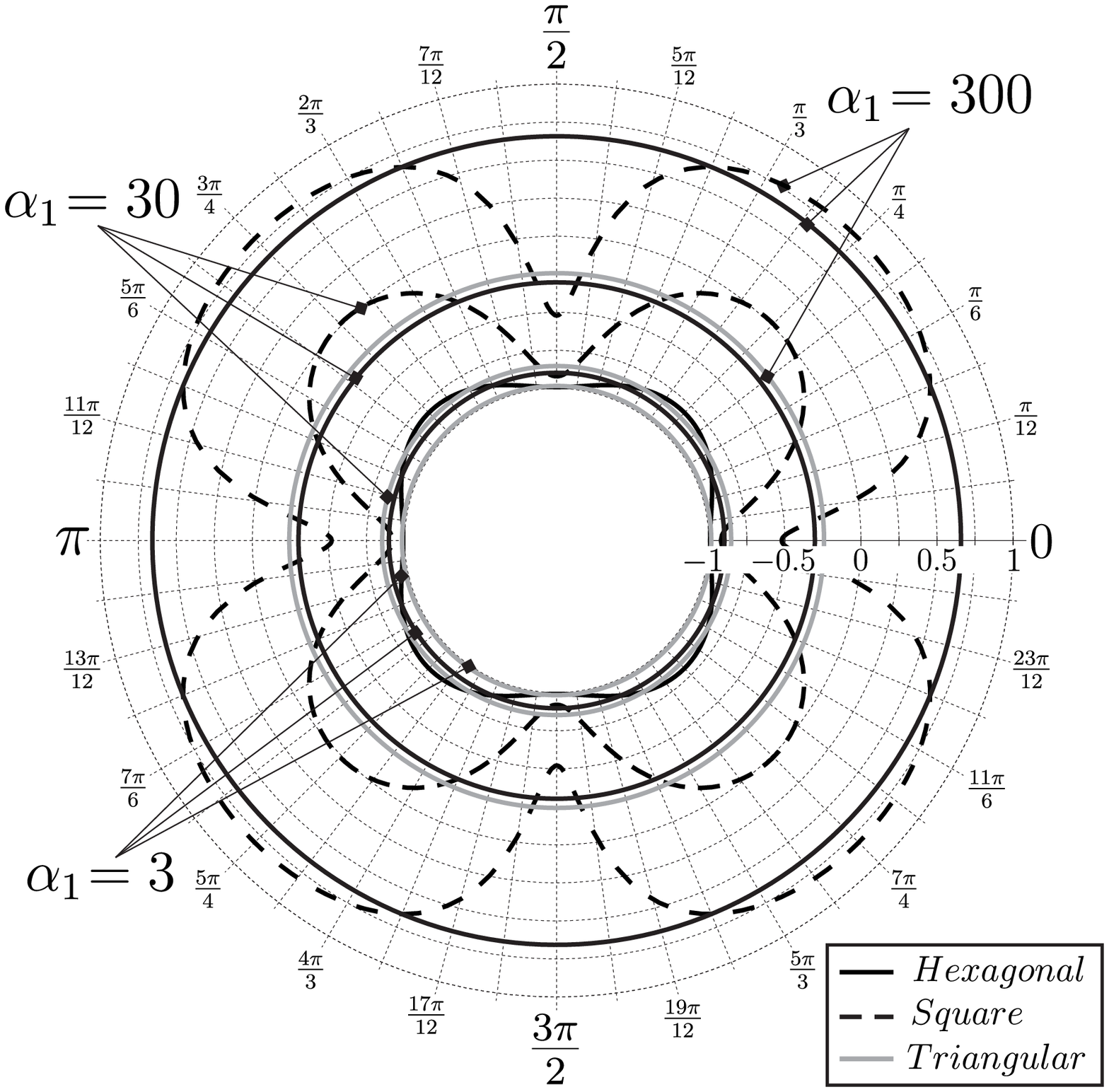} &
                \includegraphics[width=6.4 cm]{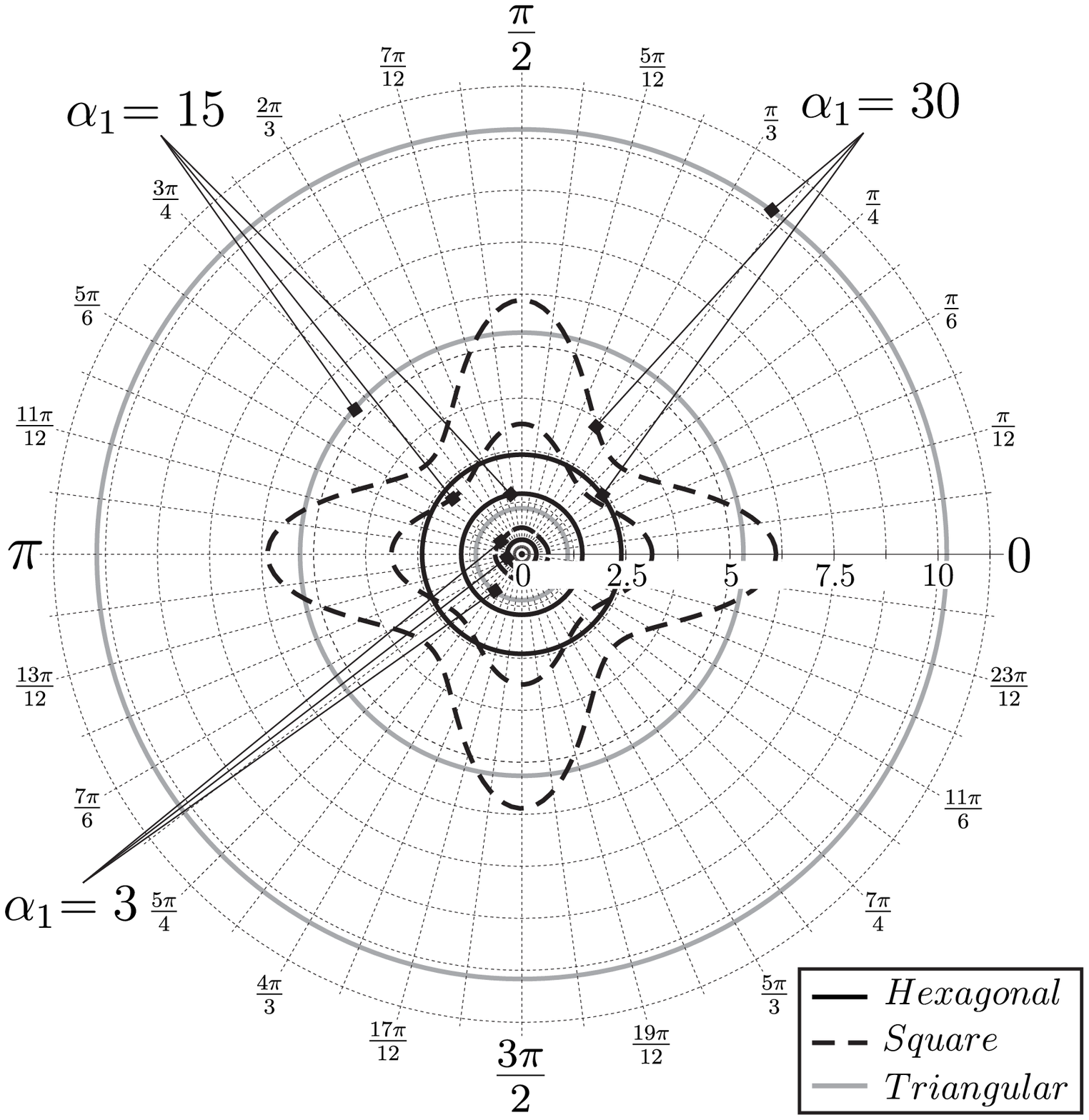}    \\
         			(a) & (b)
         \end{tabular}
}
\caption{Polar diagrams of the (a) Poisson's ratio and (b) Young's modulus. Resuls are given for the three micro-geometries with longitudinal springs corresponding to different values of the non dimensional stiffness ratio $\alpha_1=(s/p)^2\alpha_2/12$.}
\label{fig144}
\end{figure}

The polar diagrams of the Poisson's ratio and Young's modulus are given in figure \ref{fig144}. In a reference system rotated of an angle $\beta$ with respect to the system of reference $Ox_1x_2$ the Poisson's ratio and the Young's modulus are given by
\begin{equation}
\nu^*(\beta)=-\frac{b_ib_jS^*_{ijkl}n_kn_l}{n_in_jS_{ijkl}n_kn_l}, \qquad
E^*(\beta)=(n_in_jS_{ijkl}n_kn_l)^{-1}, \qquad(i,j,k,l=1,2),
\label{equ5000}
\end{equation}
where ${\bf n}=(\cos\beta,\sin\beta)^T$, ${\bf b}=(-\sin\beta,\cos\beta)^T$ and $\mathbb S^*$ is the fully-symmetric compliance tensor having components
\begin{eqnarray}
\nonumber
S^*_{1111}=S^*_{2222}=\frac{1}{E^*}, \qquad S^*_{1122}=S^*_{2211}=-\frac{\nu^*}{E^*}, \\
S^*_{1212}=S^*_{2121}=S^*_{1221}=S^*_{2112}=\frac{1}{4\mu^*},
\label{equ5001}
\end{eqnarray}
and $\mu^*=E^*/(2(1+\nu^*))$ in the isotropic cases. In addition to the verified isotropy of the hexagonal and triangular lattices, the polar plots in figure \ref{fig144} show the
increase (decrease) of the Poisson's ratio (Young's modulus) in oblique direction with a maximum (minimum) for $\beta=\pi/4$, where
\begin{eqnarray}
\nonumber
\nu^*_{S_L}(\pi/4)=-\frac{3\cos^2\gamma+3(1-\cos^2\gamma\,\sin^2\gamma)\alpha_1-\sin^4\gamma\,\alpha_2}
{3\cos^2\gamma+3(1+\cos^2\gamma\,\sin^2\gamma)\alpha_1+\sin^4\gamma\,\alpha_2}\,, \\
E^*_{S_L}(\pi/4)=\frac{6\sin^2\gamma \,k_L}
{3\cos^2\gamma+3(1+\cos^2\gamma\,\sin^2\gamma)\alpha_1+\sin^4\gamma\,\alpha_2}\,.
\label{equ5003}
\end{eqnarray}
We note that the square lattice recovers isotropy when $k_L\to 0$, in particular
\begin{equation}
\frac{\nu^*_{S_L}(\pi/4)}{\nu^*_{S_L}(0)} \simeq 1+\mathcal{O}(k_L), \qquad
\frac{E^*_{S_L}(\pi/4)}{E^*_{S_L}(0)} \simeq 1+\mathcal{O}(k_L).
\label{equ5004}
\end{equation}

\begin{figure}[!htcb]
\centerline{
         \begin{tabular}{c@{\hspace{0.5pc}}c}
                \includegraphics[width=6 cm]{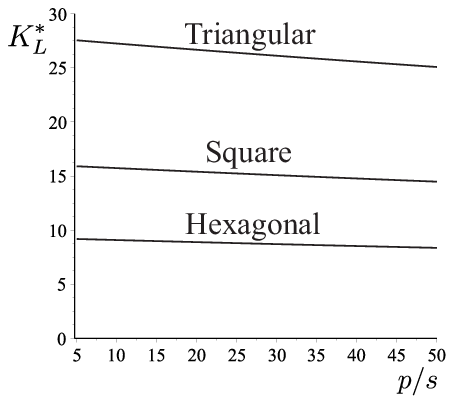} &
                \includegraphics[width=6 cm]{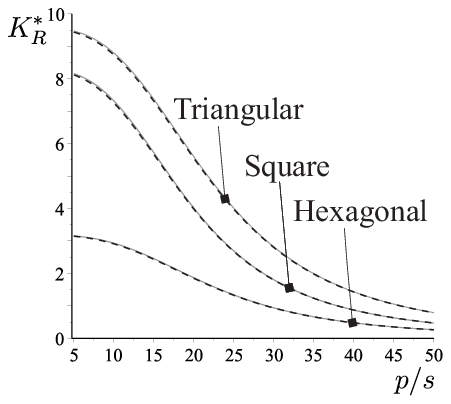}\\
         			(a) & (b)
         \end{tabular}
}
\caption{Effective bulk modulus $K^*$ [MPa] as a function of the `slenderness' $p/s$. (a) $K^*_{H_L}$, $K^*_{T_L}$ and $K^*_{S_L}$. (b) $K^*_{H_R}$, $K^*_{T_R}$ and $K^*_{S_R}$ where black dashed lines correspond to formulae in eqns. (\ref{equ211}), (\ref{equ308}) and (\ref{equ406}) while continuous grey lines correspond to the same formulae with $\alpha_3=0$. Results are given for: Young's modulus $E=3000$ MPa, $t=5$ mm, $k_L=k_R/p^2=1$ N$/$mm.}
\label{fig13}
\end{figure}

We focus now the attention on the effect of the axial and bending stiffness of the elements of the microstructure on the effective properties; 
the stiffness ratios are estimated as
\begin{equation}
\alpha_1 \simeq \frac{k_L}{E}\frac{p}{s}, \quad
\alpha_2 \simeq \frac{k_L}{E}\left(\frac{p}{s}\right)^3 \quad \mbox{and} \quad
\alpha_3 \simeq \frac{k_R/p^2}{E}\frac{p}{s}, \quad
\alpha_4 \simeq \frac{k_R/p^2}{E}\left(\frac{p}{s}\right)^3.
\label{equ503a}
\end{equation}
The ratio $p/s\gg 1$ is proportional to the slenderness of the arms of the cross-shaped elements of the microstructures. It is also inversely proportional to the density of the effective medium, as also noted in \cite{SpaRuz2011}. On physical grounds, it is reasonable to consider $\alpha_1<10$ ($\alpha_3/p^2<10$)  and, therefore $\alpha_1\ll \alpha_2^{1/3}$ ($\alpha_2\ll \alpha_4^{1/3}$).
It follows that, for sufficiently low values of $k_L$ and $k_R$, only $K^*_{H_L}$, $K^*_{T_L}$ and $K^*_{S_L}$ are governed by the axial stiffness of the arms of the cross-shaped  elements of the microstructure, while the other effective constitutive parameters are governed by  the flexural behaviour of the elements of the microstructures associated with the parameters $\alpha_2$ and $\alpha_4$. In Figure \ref{fig13} the effective bulk modulus $K^*$ is shown as a function of the geometrical parameter $p/s$; lattices with longitudinal and rotational springs are shown in parts (a) and (b), respectively. The triangular lattice is the stiffest and its bulk modulus is exactly three times the bulk modulus for the hexagonal microstructure, in fact it is easy to check that for the same $\gamma$ and the same geometrical parameters of the elements, the effective stored-energy density of the triangular lattice is exactly three times the effective stored-energy density of the hexagonal one when dilatational deformations are applied. The nearly linear dependence of $K^*_{H_L}$, $K^*_{T_L}$ and  $K^*_{S_L}$ on the `slenderness'  $p/s$ highlights the dependence of the bulk moduli on the axial stiffness of the element of the microstructure while for $K^*_{H_R}$, $K^*_{T_R}$ and  $K^*_{S_R}$, it is the bending stiffness of the elements of microstructure that governs the effective behaviour. This is evident from the comparison between the dashed black lines, corresponding to formulae in eqns. (\ref{equ211}), (\ref{equ308}) and (\ref{equ406}), and the grey continuous lines where the effect of the axial stiffness of the element has been neglected ($\alpha_3=0$).

\begin{figure}
        \centering
                \includegraphics[width=9 cm]{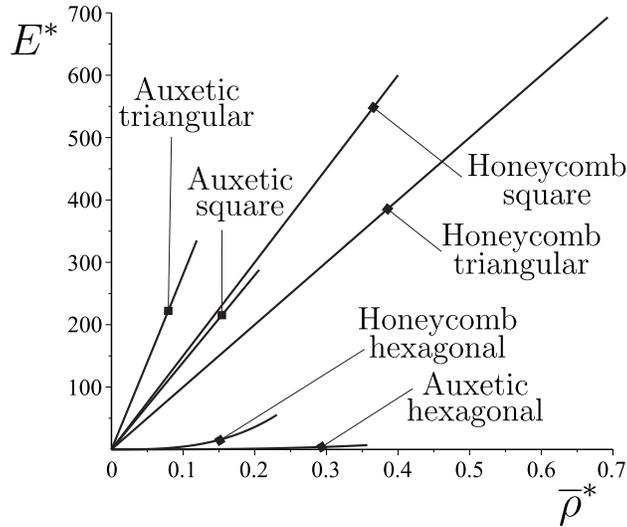}
 \caption{Comparison between auxetic and honeycomb solids. Effective Young's modulus $E^*$ [MPa] is given as a function of the relative density $\overline\rho$. The results correspond to the  parameters: Young's modulus $E=3000$ MPa (thermoplastic polymer ABS), longitudinal springs stiffness $k_L=E s/(2 p \cos\gamma)$ and $\gamma=4\pi/9$.}
\label{fig333}
\end{figure}

We conclude our analysis with a comparison of the proposed model with classical cellular solids.
Square, triangular and hexagonal lattices are common topologies encountered in physical models and engineering applications and their static behaviour is discussed in \cite{SpaRuz2011,WanMcD2004,WanMcD2005,GibAsh1997}, among others.
For the purpose of comparison we consider the behaviour in terms of effective Young's Modulus $E^*$ and we introduce the relative density parameter $\overline\rho$,
defined as the ratio between the volume occupied by the thin elements of the microstructure and the volumes of the unit cell.
In particular, for the proposed hexagonal, square and triangular auxetic lattices, we have
\begin{equation}
\overline{\rho}_{H}^*=
\frac{\sqrt{3}}{\sin^2\gamma}\frac{s}{p}\,,\quad
\overline{\rho}_{S}^*=
\frac{1}{\sin^2\gamma}\frac{s}{p}\,,\quad
\overline{\rho}_{T}^*=
\frac{1}{\sqrt{3}\sin^2\gamma}\frac{s}{p},
\label{equ504}
\end{equation}
respectively, where, for the purpose of comparison, the physical parameters are reported for unit out-of-plane thickness.

The dominant deformation mechanism in classical honeycomb structures may be of extensional or bending nature, as discussed above. The effective behaviour of honeycombs, with triangular and square microstructures, are dominated by the axial deformations of its internal components \cite{WanMcD2004}, while the corresponding hexagonal lattice is dominated by cell-wall bending \cite{GibAsh1997}. The same bending-dominated behaviour has been observed experimentally for the chiral lattice proposed in \cite{PraLak1996}.

In Figure \ref{fig333} we compare the effective Young's modulus $E^*$ of the proposed auxetic lattices with those of the honeycombs microstructures. The curves are shown for $p/s\ge 5$, in the range of validity of the beam theory, which has been used for the computation of the effective behaviour.
We note that the auxetic lattice can reach values of $E^*$ greater than the honeycomb and that the triangular topology gives the stiffest behaviour for the auxetic lattice while the square topology is the stiffest honeycomb, when the results are given in terms of relative density $\overline\rho$.

\section{Conclusions}
\label{Sect05}

A new family of auxetic lattices with Poisson's coefficient arbitrarily close to $-1$ has been proposed and the extreme properties of the micro-structured medium have been proved experimentally.
A complete analysis of the static behaviour has been performed and the effective properties have been given in closed analytical form.
They depend on the constitutive properties of the single constituents and on the topology of the microstructure.
Comparisons with classical honeycomb micro-structured media show that the parameter of the geometry can be set in order to have effective properties of the same order of magnitude.
For this type of structure we envisage quite direct applications in Structural Aeronautics and Civil Engineering, where the design of the internal hinges is a problem that has already been solved technologically, and different joint structures are already produced. Also, with the advent of 3D printing technology the capability to create the proposed microstructure at different scales open new and exciting prospects in terms of possible technological applications.

\section*{Acknowledgment}
M.B. acknowledges the financial support of the European Community's Seven Framework Programme under contract number PIEF-GA-2011-302357-DYNAMETA and of  Regione Autonoma della Sardegna (LR7 2010, grant `M4' CRP-27585). Finally, the authors would like to thank Dr. Stewart Haslinger for his valuable comments on the text.

\end{document}